\documentclass[12pt]{article}
\usepackage{a4,epsfig}

\newcommand{\be}{\begin{equation}}
\newcommand{\ee}{\end{equation}}

\newcommand{\bea}{\begin{eqnarray}}
\newcommand{\eea}{\end{eqnarray}}

\newcommand{\beq}{\begin{equation}}
\newcommand{\eeq}{\end{equation}}
\newcommand{\beqa}{\begin{eqnarray}}
\newcommand{\eeqa}{\end{eqnarray}}

\newcommand{\Eqn}[1]{Eq.~(\ref{#1})}
\newcommand{\Eqns}[2]{Eqs.~(\ref{#1}) and (\ref{#2})}

\newcommand{\bfm}[1]{\mbox{\boldmath$#1$}}

\def\bfnabla{\mbox{\boldmath $\nabla$}}

\newcommand{\RS}{\rm RS}

\newcommand{\MS}{\overline{\rm MS}}
\def\als{\alpha_{\rm s}}
\newcommand{\nn}{\nonumber}

\begin{document}

\begin{flushright}
  IPPP/10/41 \\
  KEK-TH-1368\\
\end{flushright}

\vspace*{1cm}

\begin{center}
  {\sc \large Improved determination of inclusive electromagnetic\\[5pt]
    decay ratios of heavy quarkonium from qcd} \\
   \vspace*{2cm} {\bf Yuichiro~Kiyo$^a$, Antonio~Pineda$^b$
and Adrian~Signer$^c$}\\
\vspace{0.6cm}
{\it $^a$\ Theory Center, KEK, Tsukuba, Ibaraki 305-0801, Japan
\\[10pt]
$^b$\ Grup de F\'\i sica Te\`orica and IFAE, Universitat
Aut\`onoma de Barcelona, E-08193 Bellaterra, Barcelona, Spain
\\[10pt]
$^c$\ Institute for Particle Physics Phenomenology \\
Durham, DH1 3LE, England \\}
  \vspace*{2.4cm}
  {\bf Abstract} \\
  \end{center}
  We consider a different power counting in potential NRQCD by
  incorporating the static potential exactly in the leading order
  Hamiltonian. We compute the leading relativistic corrections to the
  inclusive electromagnetic decay ratios in this new scheme. The
  effect of this new power counting is found to be large (even for
  top). We produce an updated value for the $\eta_b$ decay to two
  photons. This scheme also brings consistency between the weak
  coupling computation and the experimental value of the charmonium
  decay ratio.

\thispagestyle{empty}

\vspace*{5mm}
\noindent

\newpage

\setcounter{page}{1}
\pagestyle{plain}



\section{Introduction}

The determination of heavy quarkonium properties from QCD has always
been a major objective in high energy physics. In this respect, the
development of effective field theories (EFT) directly derived from
QCD like NRQCD~\cite{Caswell:1985ui} or pNRQCD~\cite{Pineda:1997bj}
(for a review see Ref.~\cite{Brambilla:2004jw}) has opened the door to
model independent determinations of heavy quarkonium properties.
Instrumental in this development is the fact that heavy quarkonium
systems can be considered to be non-relativistic (NR). They are then
characterized by, at least, three widely separated scales: hard (the
mass $m$, of the heavy quarks), soft (the relative momentum $|{\bf p}|
\sim mv \ll m$, of the heavy-quark--antiquark pair in the center
of mass frame), and ultrasoft (the typical kinetic energy $E \sim
mv^2$ of the heavy quark in the bound state system).

In this paper we focus on pNRQCD. This EFT takes full advantage of the
hierarchy of scales that appear in the system,
\begin{equation}
\label{hierarchy}
m \gg mv \gg mv^2 \cdots
\,,
\end{equation}
and makes a systematic and natural connection between quantum
field theory and the Schr\"odinger equation. Schematically the EFT
takes the form
\begin{eqnarray*}
\,\left.
\begin{array}{ll}
&
\displaystyle{
\left(i\partial_0-{{\bf p}^2 \over m}-V_s^{(0)}(r)\right)\Phi({\bf r})=0}
\\
&
\displaystyle{\ + \ \mbox{corrections to the potential}}
\\
&
\displaystyle{\ +\ 
\mbox{interactions with other low-energy degrees of freedom}}
\end{array} \right\}
{\rm pNRQCD}
\end{eqnarray*}
where $V_s^{(0)}(r)$ is the static potential and $\Phi({\bf r})$ is
the $Q$-$\bar{Q}$ wave function.

A major issue to be settled is to decide upon the precise form of
$V_s^{(0)}(r)$, in particular whether one works in the weak or strong
coupling regime and how to treat subleading terms. In the strict weak
coupling regime one could approximate the static potential by the
Coulomb potential $V_s^{(0)}(r)\simeq V_C = -C_F\, \als/r$ and include
higher-order terms perturbatively. There seems to be growing consensus
that the weak coupling regime appears to work properly for $t$-$\bar
t$ production near threshold, the bottomonium ground state mass, and
bottomonium sum rules (for a recent discussion on this issue see
\cite{Pineda:2009zz}). One would then expect that other properties of
the bottomonium ground state like the hyperfine splitting or
electromagnetic decay widths could be described as well by the weak
coupling version of pNRQCD. However, in this case the situation is not
that clear.  There has been a precise determination of the bottomonium
ground state hyperfine splitting using the renormalization group in
pNRQCD \cite{Kniehl:2003ap}. Nevertheless, the predicted value does
not agree well with the recently obtained experimental number
\cite{:2008vj,Bonvicini:2009hs}. Therefore the situation remains
unsettled. For the inclusive electromagnetic decays the convergence is
not very good \cite{Pineda:2006ri}. Even for top, higher-order
corrections to the normalization appear to be
sizable~\cite{Pineda:2006ri, Hoang:2003ns, Beneke:2005hg,
  Beneke:2007gj}.

In principle, the novel feature of these observables (maybe more so
for the decays) compared to the heavy quarkonium ground state mass is a
bigger sensitivity to the value of the wave function at the origin and
to its relativistic corrections. Note that in this case the
relativistic corrections are divergent and their divergences have to
be absorbed by the matching coefficients of the effective theory:
potentials and current matching coefficients. If one considers the
decay ratio, the dependence on the wave function associated to the
static potential drops out and only the relativistic correction
survives. This makes the decay ratio the cleanest possible place on
which to quantify the importance of the relativistic corrections to
the wave function.

In Ref. \cite{Penin:2004ay} the decay ratios have been computed with
NNLL accuracy, accounting for the resummation of logarithms. The scale
dependence has greatly improved over fixed-order computations and the
result is much more stable.  The convergence could be classified as
good for the top case, reasonable for the bottom, and not good for the
charm, although in all three cases the scale dependence of the
theoretical result was quite small. For the case of the charm there is
experimental data available, and the agreement with experiment
deteriorates when higher order corrections are introduced.  On the
other hand there exists a nice analysis for charmonium in
Ref. \cite{Czarnecki:2001zc}, where they consider a potential model (a
Cornell-like one, yet compatible with perturbation theory at short
distances, since it is coulomb-like in this regime) for the bound
state dynamics, but a tree-level perturbative potential for the
spin-dependence. They also correctly performed the matching in the
ultraviolet with QCD along the lines of what would be pNRQCD in the
strong coupling regime\footnote{Actually the whole computation would
  fit into the strong coupling regime of pNRQCD except for the fact
  that the spin-dependent potential is computed in perturbation
  theory.}. Their net result was that they were able to obtain
consistency with experiment albeit with large errors. Unfortunately,
this result suffers from model dependence. In particular, since a
perturbative potential has been used for the spin-dependent potential,
it would have been more consistent to treat the static potential also
in a perturbative approach. In this respect, it has been shown in
Refs. \cite{Recksiegel:2001xq,Pineda:2002se,Lee:2002sn,Brambilla:2009bi}
that, once the renormalon cancellation is taken into account, the
inclusion of perturbative corrections to the static potential leads to
a convergent series and that this series gets closer to the lattice
values in the quenched approximation up to scales of around 1 GeV. It
is then natural to ask whether the inclusion of these effects may lead
to a better agreement in the case of charmonium and for sizable
corrections in the case of bottomonium and $t$-$\bar t$ production
near threshold.  Note that in this
comparison between lattice and perturbation theory one has to go to
high orders to get good agreement. Therefore, a computation of the
relativistic correction based on the leading order expression for the
static potential, i.e.  the Coulomb potential, as the one used in an
strict NNLL computation, may lead to large corrections, since these
corrections, as well as the wave function at the origin, could be
particularly sensitive to the shape of the potential.

Therefore, in this paper we reorganize the perturbative expansion and
consider the static potential exactly, whereas we treat the the
relativistic terms as corrections. By doing so we expect to have an
effect similar to the one observed in Ref.~\cite{Czarnecki:2001zc}.
Including also the renormalization group improved expressions, we
expect to obtain results with only a modest scale dependence.  The
explicit computation will confirm to a large extent these
expectations. We will be able to give an updated prediction for the
decay of the $\eta_b$ to two photons and obtain a result for the charm
decay ratio compatible with experiment (though in this last case with
rather large errors).  Note that our computation is
completely based on a weak coupling analysis derived from QCD 
and no non-perturbative input is introduced.

\section{Decay ratio}

The one-photon mediated processes are induced by the electromagnetic
current $j_\mu$, which has the following decomposition in
terms of operators constructed from the non-relativistic quark and
anti-quark two-component Pauli spinors $\psi$ and $\chi$ \cite{BBL}:
\begin{equation}
\bfm{j}=c_v(\mu)\psi^\dagger{\bfm\sigma}\chi+{d_v(\mu)\over6m_q^2}
\psi^\dagger\bfm{\sigma}\mbox{\boldmath$D$}^2\chi
+\ldots,
\label{vcurr}
\end{equation}
where $\mu$ is the
renormalization scale, $\bfm{D}$ is the covariant
derivative, ${\bfm\sigma}$ is the Pauli matrix, and the ellipsis stands
for operators of higher mass dimension.  The Wilson coefficients $c_v$
and $d_v$ represent the contributions from the hard modes and may be
evaluated as a series in $\alpha_s$ in
full QCD for free on-shell on-threshold external (anti)quark fields.
We define it through
\begin{eqnarray}
  c_v(\mu) &=& \sum_{i=0}^\infty\left(\alpha_s(\mu)\over
  \pi\right)^i c_v^{(i)}(\mu)\
  \,, \qquad c_v^{(0)}=1\,,
\end{eqnarray}
and similarly for other coefficients. The coefficients $c_v^{(1)}$ and
$c_v^{(2)}$ have been computed in Refs.~\cite{KalSar} and
\cite{CzaMel1,BSS} respectively.

The operator responsible for the two-photon $S$-wave  processes in the
non-relativistic limit is generated by the expansion of the product of
two electromagnetic currents and has the following representation~\cite{BBL}
\begin{equation}
O_{\gamma\gamma}=c_{\gamma\gamma}(\mu)\psi^\dagger\chi
+{d_{\gamma\gamma}(\mu)\over6m_q^2}
\psi^\dagger\mbox{\boldmath$D$}^2\chi
+\ldots,
\label{gcurr}
\end{equation}
which reduces to the pseudo-scalar current in the non-relativistic
limit.  The coefficients $c_{\gamma\gamma}^{(1)}$ and
$c_{\gamma\gamma}^{(2)}$ have been computed in Refs.~\cite{HarBro} and
\cite{CzaMel2} (in semi-numerical form) respectively.

Let us define the spin ratio for the production and annihilation of
heavy quarkonium ${\cal Q}$ as
\begin{equation}
{\cal R}_q=
{\sigma(e^+e^-  \rightarrow {\cal Q}(n^3S_1) )\over
\sigma(\gamma\gamma \rightarrow {\cal Q}(n^1S_0))}=
{\Gamma({\cal Q}(n^3S_1)\to e^+e^-)\over
\Gamma({\cal Q}(n^1S_0)\to  \gamma\gamma)}\,.
\end{equation}
The effective theory expression for the spin ratio reads
\begin{equation}
{\cal R}_q={c_s^{\,2}(\mu)\over 3Q_q^2}
{|\psi_n^{v}(0)|^2\over|\psi_n^{p}(0)|^2}+{\cal O}(\alpha_s v^2)\,,
\label{Rdef}
\end{equation}
where $Q_q$ is the quark electric charge,
$c_s(\mu)=c_v(\mu)/c_{\gamma\gamma}(\mu)$, $\psi_n^{(v,p)}(\bfm{r})$
are the spin triplet (vector) and spin singlet (pseudo-scalar)
quarkonium wave functions with principal quantum number $n$.  The wave
functions describe the dynamics of the non-relativistic bound state
and can be computed within pNRQCD.  The latter is the
Schr\"odinger-like effective theory of potential (anti)quarks whose
energies scale like $m_qv^2$ and three-momenta scale like $m_qv$, and
their multipole interaction to the ultrasoft gluons~\cite{KniPen1,
  BPSV2, Beneke:2007pj, Beneke:2008cr}.  The contributions of hard and
soft modes in pNRQCD are represented by the perturbative and
relativistic corrections to the effective Hamiltonian, which is
systematically evaluated order by order in $\alpha_s$ and $v$ around
the leading order (LO) Coulomb approximation.

\section{pNRQCD framework}

As we have mentioned before, the framework we use to compute the decay
ratio, and more specifically the wave function, is pNRQCD. For the
purposes of our paper the full setup of pNRQCD is not needed. We will
only need the static potential, $V_s^{(0)}(r)$, and the spin-dependent
potential ${V}^{(2)}_{S^2,s}(r)$. Furthermore, we will
reorganize the perturbative expansion. The static potential will be
treated exactly by including it in the leading-order Hamiltonian
\begin{eqnarray}
\label{H0}
H^{(0)}\equiv -\frac{{\bf \nabla}^2}{2m_r}+V^{(0)}_s(r),
\end{eqnarray}
where $m_r=m_1m_2/(m_1+m_2)$. On the other hand, the spin-dependent
potential (in $D= 1+d= 4-2\epsilon$ dimensions)
\begin{equation}
\label{DeltaH}
\Delta H=
\frac{V^{(2)}_{S^2,s}(\mu)}{m_1m_2}=
- \frac{4\pi C_F D^{(2)}_{S^2,s}}{d\, m_1m_2}\,
   [{\bf S}_1^i,{\bf S}_1^j][{\bf S}_2^i,{\bf S}_2^j]
   \delta^{(d)}({\bf r})
\end{equation}
is considered to be a perturbation to the result obtained with
$H^{(0)}$.  Therefore, we distinguish between an expansion in $v$ and
$\als$. $v$ has an expansion in $\als$ itself but this expansion does
not converge quickly for these relativistic corrections. This remains
so even after the inclusion of the renormalon cancellation,
which has only a minor impact on the determination of the wave
function. This is the reason we choose to take the static potential
exactly. 

As mentioned in the introduction there are different options on how
precisely to treat $V_s^{(0)}$ and we will discuss in
Section~\ref{sec:BrVs} the various options we consider. Roughly
speaking we will take the static potential up to NNNLO including also
the leading ultrasoft corrections. We will also need to define a scheme of
renormalon subtraction.  Therefore, the general form of the static
potential will be
\begin{equation}
\label{VsRen}
V_s^{(0)}(r)=V_{SD}(r)+2\, \delta m_X
\,,
\end{equation}
where $\delta m_X$ represents a residual mass that encodes the pole
mass renormalon contribution and $X$ stands for the specific
renormalon subtraction scheme. We will show some specific examples in
Section~\ref{sec:RGIpot}. In \Eqn{VsRen}, $V_{SD}$ is the short
distance behavior of the static potential, which is independent of the
scheme for renormalon subtraction (even if we use a non-perturbative
potential).  In momentum space it reads
\begin{equation}
\lim_{q \rightarrow \infty} \widetilde{V}^{(0)}_s(q)=\widetilde{V}_{SD}(q)
=
-\frac{4\pi C_F\,
\widetilde{\alpha}_{V^{(0)}_s}(q)}{{\bf q}^2},
\label{eq:StaticPotential}
\end{equation}
with $\widetilde{\alpha}_{V^{(0)}_s}(q) \sim \alpha_s(\mu)$ (for the
precise relation see \Eqn{eq:VtildeSDfo}), where $\alpha_s(\mu)$ is the
QCD coupling constant in the $\overline{\rm MS}$-scheme.

For the spin-dependent potential in momentum space we have
\begin{eqnarray}
\widetilde{V}^{(2)}_{S^2}(\mu)
&=&
- \frac{4\pi C_F D^{(2)}_{S^2,s}(\mu)}{d\,}\,
       [{\bf S}_1^i,{\bf S}_1^j][{\bf S}_2^i,{\bf S}_2^j]
\nonumber \\
&=&
- \frac{4\pi C_F D^{(2)}_{S^2,s}(\mu)}{3}\,
   \left(\frac{3}{2}-S^{\,2}
         +\epsilon\left(\frac{9}{2}-\frac{8}{3}S^{\,2}\right)
   \right)+ {\cal O}(\epsilon^2)\,,
\label{eq:spin_dep_pot}
\end{eqnarray}	
where ${\bf S}_{1,2}$ is the spin operator for heavy quark and
anti-quark, respectively and
$D_{S^2,s}^{(2)}(\mu)=\alpha_s(\mu)+\ldots$.  In the second line in
Eq.(\ref{eq:spin_dep_pot}) the spin projection has been done,
resulting in $S^2\equiv 0$ and 2 for spin-singlet and spin-triplet
states, respectively (this expression actually corresponds to the
regularization prescription of \cite{Czarnecki:2001zc} for the
spin-zero states).  We have to keep the term of ${\cal O}(\epsilon)$
because the spin-dependent potential generates $1/\epsilon$
divergences.  The renormalization procedure for these $1/\epsilon$
will be discussed in the next section.

\section{Wave function ratio}

We now turn to the computation of
\begin{eqnarray}
\label{rho_n}
  \frac{|\psi_n^{v}(0)|^2}{|\psi_n^{p}(0)|^2} &\equiv&
  \rho_n(\mu)\,\,\equiv\,\,1+\delta \rho_n(\mu)
  \,,
\end{eqnarray}
Applying Rayleigh-Schr\"odinger perturbation theory to the problem we
obtain
\begin{eqnarray}
\psi^{v/p}_n(0)
&=&
\psi_{n}^{(0)}(0)
-
\widehat{G}(E^{(0)}_n)
\frac{\widetilde{V}^{(2)}_{S^2}(\mu)}{m_1m_2}\, \psi_n^{(0)}(0)\,
+ {\cal O}\left(\tilde V_{S^2}^{(2)}\right)^2, \,
\label{eq:psi_at_0}
\end{eqnarray}
where $\psi_n^{(0)}(0)$ is the wave function for the LO Hamiltonian
$H^{(0)}$ and $\widehat{G}(E^{(0)}_n)$ is the reduced Green function
at $E=E^{(0)}_n$, which is defined by
\begin{equation}
\widehat{G}(E^{(0)}_n)
\equiv
\sum_m{}^{\prime}
\frac{|\psi^{(0)}_m(0)|^2}{E^{(0)}_m-E^{(0)}_n}
=
\lim_{E\rightarrow E^{(0)}_n}
\bigg(
G(E)-\frac{|\psi^{(0)}_n(0)|^2}{E^{(0)}_n-E}
\bigg)\, .
\label{eq:Gred}
\end{equation}
The prime indicates that the sum does not include the state $n$ and
\begin{equation}
G(E) = G(0,0;E)
\equiv
\lim_{r \rightarrow 0}G(r,r;E)
=
\lim_{r \rightarrow 0}\langle {\bf r}| \frac{1}{H^{(0)}-E-i 0}|{\bf r} \rangle
\,
\end{equation}
is the zero-distance limit of the Green function $G(r,r';E)$, which is
the solution of the Schr\"odinger equation
\begin{eqnarray}
&&
\bigg[-\frac{\bfnabla^2}{2m_r}+V_s^{(0)}(r)-E \bigg] G(r, r'; E)
=\delta({\bf  r}-{\bf r}').
\end{eqnarray}
The short distance behavior of the static potential $V_s^{(0)}(r)\sim
1/r$ makes $G(E)$ and, therefore, $\delta\rho_n$ divergent. Thus we
will need to regularize the Green function and we will deal with two
different ways to do this: dimensional regularization and finite-$r$
regularization. We start by considering the former and will come back
to finite-$r$ regularization in the next section.

The divergences in $\delta \rho_n$ are cancelled by divergences in the
Wilson coefficient $c^2_s(\mu)$. Since the latter have been computed
in dimensional regularization we will need $G(E)$ in dimensional
regularization as well. We denote the corresponding bare and reduced
Green functions by $G^{(D)}(E) =G^{(D)}(0,0;E)$ and
$\widehat{G}^{(D)}(E^{(0)}_n)$ respectively. We remark that the LO wave
functions (corresponding to $H^{(0)}$) are finite, thus
$|\psi^{(0)(D)}_n(0)|^2=|\psi^{(0)(4)}_n(0)|^2\equiv
|\psi_n^{(0)}(0)|^2$.

Using Eqs.~(\ref{eq:spin_dep_pot})--(\ref{eq:psi_at_0}), the bare
expression of $\delta \rho_n(\mu)$ in dimensional regularization can
be written as
\begin{eqnarray}
&&
\delta \rho_n^{(D)}(\mu)
=
-\frac{16\pi C_F}{3 m_1m_2}
D_{S^2,s}^{(2)}(\mu)
\left(1 +\frac{8}{3}\,\epsilon+{\cal O}(\epsilon^2)\right)
 \widehat{G}^{(D)}(E^{(0)}_n).
\label{eq:rho_formula}
\end{eqnarray}
In order to obtain the $\MS$-renormalized expression of $\delta
\rho_n$, we need to identify the divergences of
$\widehat{G}^{(D)}(E^{(0)}_n)$. They are the same as those of $G^{(D)}(E)$,
are independent of $E$, and can be computed order by order in
perturbation theory, since they are related to the short distance
behavior of the Green function. We thus parameterize the divergent and
finite terms of $G^{(D)}(E)$ and $\widehat{G}^{(D)}(E^{(0)}_n)$ as
\begin{eqnarray}
G^{(D)}(E)&=&\frac{m_r}{2\pi}
\bigg[A_{\MS}^{(D)}(\epsilon;\mu)+B_{V_s^{(0)}}^{\MS}(E;\mu)\bigg]\,,
\label{eq:GD}
\\
\widehat{G}^{(D)}(E^{(0)}_n)&=&\frac{m_r}{2\pi}
\bigg[A_{\MS}^{(D)}(\epsilon;\mu)+\widehat{B}_{V_s^{(0)}}^{\MS}(E^{(0)}_n;\mu)\bigg]\,,
\label{eq:GDhat}
\end{eqnarray}
where $B_{V^{(0)}_s}^{\MS}(E;\mu)$ and
$\widehat{B}_{V^{(0)}_s}^{\MS}(E^{(0)}_n;\mu)$ are finite in 4 dimensions,
but contain terms to all orders in $\alpha_s/v$, since the bound-state
dynamics needs all order resummation in $\alpha_s$. As will be shown,
\Eqn{GCbare}, the ultraviolet divergent part can be expressed in terms
of the (dimensionfull) bare coupling $g^2\equiv 4\pi \als\,
\mu^{2\epsilon}$ as
\begin{equation}
A^{(D)}_{\MS}(\epsilon;\mu)
=\frac{g^2\, C_F\, m_r}{8\pi\epsilon}
\left(\frac{\mu^2 e^{\gamma_E}}{4\pi}\right)^{-2\epsilon}+{\cal O}(\alpha_s^2)
\, .
\label{eq:AD}
\end{equation}
$A_{\MS}^{(D)}$ will be removed by renormalization.  This has to be
done consistently with the calculation of other parts order by order
in the expansion in $\alpha_s$ (in our case $\MS$).  The divergences
are then absorbed in $c_s$ and we can write
\begin{eqnarray}
&&
\delta \rho_n^{\MS}(\mu)
=
-\frac{8 m_r C_F}{3 m_1m_2}
D_{S^2,s}^{(2)}(\mu)
\left(\widehat{B}_{V^{(0)}_s}^{\MS}(E^{(0)}_n;\mu) +
\frac{4}{3}m_rC_F\als+{\cal O}(\als^2)\right).
\label{eq:rho_formulaMS}
\end{eqnarray}
This will have to be combined with the $\MS$ subtracted matching
coefficient $c_s^2(\mu)$ in \Eqn{Rdef} to obtain the decay ratio.
 
\section{Green Function in position space}
\label{Gr}

The main goal of the present paper is to compute
$\widehat{G}^{(D)}(E^{(0)}_n)$ or, equivalently,
$\widehat{B}_{V^{(0)}_s}^{\MS}(E^{(0)}_n;\mu)$, with the effect of the
static potential included exactly. This calls for a numerical
evaluation of the Green function rather than pursuing an analytic
approach.  Numerical calculations are most conveniently performed in
coordinate space. It is here where finite-$r$ regularization comes
into play. In Section~\ref{Gr:reg} we will discuss this regularization
and in Section~\ref{Gr:match} we show how to convert the Green
function obtained in finite-$r$ regularization by matching into the
one in dimensional regularization.

\subsection{Regularization of the Green function in position space}
\label{Gr:reg}

The zero-distance Green function with finite-$r$ regularization is
simply defined as $G^{(r)}(E)\equiv G(r_0, r_0; E)$, where $r_0\ll
1/(m\alpha_s)$. In order to compute it, we first have to describe how
to obtain a numerical solution for the Green function $G(r,r';E)$ in
general, given the static potential $V^{(0)}_s(r)$.  Actually the
whole procedure holds valid for a generic potential (not unbounded
from below at long distances) that has the correct, perturbative,
short distance limit\footnote{This opens the possibility of using the
  same formalism for pNRQCD in the strong coupling regime but then we
  should also consider a non-perturbative potential in
  Eq.~(\ref{DeltaH}), albeit with the correct short distance limit.}.
According to \Eqn{VsRen} renormalon associated affects are power
suppressed.  Therefore, they will not affect properties associated to
the $r \rightarrow 0$ limit of the potential.

For the class of potentials described above, the Green function
$G(r,r'; E)$ can be constructed from the two independent solutions
$u_<(r), u_>(r)$ of the homogeneous Schr\"odinger equation (our
approach follows Ref.~\cite{Strassler:1990nw}, see also
Ref.~\cite{Melnikov:1998pr})
\begin{eqnarray}
\left[\frac{d^2}{dr^2} + 2m_r \left(E-V_s^{(0)}(r)\right)\right]\,u(r)=0.
\label{eq:diff_eq_for_u}
\end{eqnarray}
Here $u(r)$ represents $u_<(r)$ or $u_>(r')$, which are the solutions
to \Eqn{eq:diff_eq_for_u} that are regular for $r\rightarrow 0$ and
$r'\rightarrow \infty$ respectively.  The angular-momentum term is
dropped in the Schr\"odinger equation assuming S-wave contribution
because the limit $r,r'\rightarrow 0$ is taken later. The Green
function is written as
\begin{eqnarray}
G(r,r'; E)=\left(\frac{m_r}{2\pi}\right)\,\frac{u_<(r)}{r}\,\frac{u_>(r')}{r'}
\hspace{1cm} \mbox{for}~~ r < r'.
\end{eqnarray}

The numerical solution at finite $r$ is obtained by solving the
Schr\"odinger equation with boundary conditions at short distances.
To this end we prepare two independent solutions $u_{0}(r)$ and
$u_{1}(r)$ that are defined by the following initial conditions: For
$u_1(r)$, which we will call the regular solution we set
\begin{equation}
u_1(0)=0 \qquad {\rm and} \qquad u_1^{\prime}(0)=1
\end{equation}
so that
\begin{equation}
u_1(r)=  r + {\cal O}(r^2).
\end{equation}
This completely fixes $u_1(r)$.

For the non-regular solution, $u_0(r)$, we can not work this
way. Whereas we can still take $u_0(0)=1$, we can not define
$u_0^{\prime}(0)$, as it becomes singular.  Therefore, we first define
$u_0^{\prime}$ for small values of $r$ in the following way
\begin{eqnarray}
u'_0(r)&=&C_0(r_c)+2m_r\,\int_{r_c}^r\,dr'\, V_{SD}(r')+{\cal O}(r),
\end{eqnarray}
where $C_0(r_c)$ is an integration constant.  Note that $r_c > 0$ acts
as a cutoff to avoid the denominator-zero of $V_{SD}(r')$.  The total
solution then reads (at short distances)
\begin{equation}
u_0(r)=1+C_0(r_c)\, r+2m_r\int_0^rdr^{\prime}
\int_{r_c}^{r^{\prime}}dr^{\prime\prime} V_{SD}(r^{\prime\prime})+{\cal O}(r^2)
\,.
\end{equation}
This expression can be rewritten as
\begin{equation}
u_0(r) = 1 + C_0(r_c)\, r +
2m_r r \bigg\{ \int_{r_c}^{r}dr' V_{SD}(r')
           - \int_{0}^{r} dr'  \frac{r' V_{SD}(r')}{r}
     \bigg\} +{\cal O}(r^2).
\label{eq:bc_for_u}
\end{equation}
The derivatives $u'_{0,1}(r)$ and $u_{0,1}(r)$ at small $r$
are used as boundary conditions to solve differential equations
by, for instance, the Runge-Kutta method.
For later convenience we take $r_c=1/(\mu e^{\gamma_E})$ and fix 
\begin{eqnarray}
C_0(r_c)
=
-\frac{2m_r}{r_c}\int_0^{r_c}dr'\int_{r_c}^{r'}dr^{\prime\prime}
V_{SD}(r^{\prime\prime})
=
2m_r\int_{0}^{r_c} dr'\,  \frac{r'  V_{SD}(r')}{r_c}\,.
\label{eq:C0}
\end{eqnarray}
With this choice the ${\cal O}(r)$ term of \Eqn{eq:bc_for_u} for $u_0$
is a function of $\ln\left(\mu e^{\gamma_E} r\right)$ with no
log-independent terms
\begin{equation}
u_0(r)
=
1+2m_r r\sum_{n=1}^{\infty}v_n\ln^n(\mu e^{\gamma_E}r)+{\cal O}(r^2)
\,.
\end{equation}
The coefficients $v_n$, which can be written as an expansion in powers
of $\als(\mu)$, only depend on the coefficients $a_n$ of $V_{SD}$ (see
\Eqn{VSDfo}), i.e.  only on the pure short distance behavior of the
static potential.  This choice will turn out to be very convenient for
the conversion to dimensional regularization, but the final result for
$G(r,r;E)$ does not depend on this choice.

From the two solutions $u_{0}(r)$ and $u_{1}(r)$ we can construct
$u_>(r)$ and $u_<(r)$ as follows: First the solution at short distance
$u_<(r)$ is identified as
\begin{eqnarray}
u_<(r) &=& u_1(r)\,,
\end{eqnarray}
because $\lim_{r\rightarrow 0} u_1(r)=0$.  The other solution which
satisfies $\lim_{r\rightarrow \infty} u_>(r)=0$ is given by
\begin{eqnarray}
u_>(r) &=&
u_0(r) + B^{(r)}_{V_s^{(0)}}(E)\, u_1(r)\,,
\\
B^{(r)}_{V_s^{(0)}}(E)&=&-\lim_{r\rightarrow \infty}\,
\left\{\,u_0(r)/u_1(r)\,\right\}\,.
\end{eqnarray}
From the boundary conditions of $u_0$ and $u_1$ it follows that we can
mix a $u_1$-component into $u_0(r)$. However, the precise choice of
$u_0(r)$ does not affect $u_>(r)$ because of the invariance under
$u_0(r)\rightarrow u_0(r)+\kappa\, u_1(r)$ with $\kappa$ being an
arbitrary constant.  The zero-distance Green function with finite-$r$
regularization is then obtained as
\begin{eqnarray}
G^{(r)}(E)
&=&
\frac{m_r}{2\pi}
\bigg[ A^{(r)}(r_0;\mu)+B_{V_s^{(0)}}^{(r)}(E;\mu)\bigg]\, ,
\label{eq:Gr}
\\
A^{(r)}(r_0;\mu)
&=&
\frac{u_0(r_0)}{r_0} 
=
\frac{1}{r_0} -2m_rC_F\alpha_s \ln\left(\mu\, e^{\gamma_E}r_0\right)
 +{\cal O}(\alpha_s^2)\,,
\label{eq:Ar}
\end{eqnarray}
where the last equality is a good approximation for $\mu
e^{\gamma_E}r_0 \sim 1$. $A^{(r)}(r_0;\mu)$ encodes the divergence of
$G^{(r)}(E)$ and plays the role of the $1/\epsilon$ pole of
$G^{(D)}(E)$. It is energy independent because it is related to the
overall divergence of the Green function.  Nevertheless, note that
according to \Eqn{eq:Gr} we define $B_{V_s^{(0)}}^{(r)}(E;\mu)$ by
subtracting exactly $u_0(r_0)/r_0$, which, depending on the potential,
will include terms with arbitrary powers of $\als$.  $B^{(r)}(E;\mu)$
is computed numerically by solving the Schr\"odinger equation and is
independent of the regulator $r_0$. Note however that it is scheme
dependent, i.e. it depends on the specific condition we use for
$u_0'(r_0)$. This dependence cancels between $A^{(r)}$ and
$B_{V_s^{(0)}}^{(r)}$ such that $G^{(r)}(E)$ is independent of the
specific choice for $u_0'(r_0)$.  In analogy to \Eqn{eq:GDhat} we also
define
\begin{equation}
\widehat{G}^{(r)}(E^{(0)}_n)=\frac{m_r}{2\pi}
\bigg[A^{(r)}(r_0;\mu)+\widehat{B}_{V_s^{(0)}}^{(r)}(E^{(0)}_n;\mu) \bigg]\, .
\label{eq:Grhat}
\end{equation}

Finally we remark that $B_{V_s^{(0)}}^{(r)}$ is independent of the
renormalon subtraction scheme used, since $\widehat{G}^{(r)}(E^{(0)}_n)$ and 
$A^{(r)}(r_0;\mu)$ are; the latter by the definition used in this paper. 

\subsection{Conversion to the $\MS$ scheme}
\label{Gr:match}

Once we have the zero-distance Green function $G^{(r)}(E)\equiv G(r_0,
r_0; E)$, where $r_0\ll 1/(m\alpha_s)$, or more precisely
$B_{V_s^{(0)}}^{(r)}$, we have to convert the result by matching into
the one in dimensional regularization $B_{V^{(0)}_s}^{\MS}$, in order
to be able to use \Eqn{eq:rho_formulaMS}. We define the difference
\begin{equation}
c^{\MS}_r=
B^{\MS}_{V_s^{(0)}}(E;\mu) - B^{(r)}_{V_s^{(0)}}(E;\mu) = 
\widehat{B}^{\MS}_{V_s^{(0)}}(E^{(0)}_n;\mu) - 
\widehat{B}^{(r)}_{V_s^{(0)}}(E^{(0)}_n;\mu) \, .
\label{eq:BD_Br}
\end{equation}
This difference between the schemes can be accounted for by a finite
($r_0$ and $\epsilon$ independent) constant.  We also use the fact
that the ultraviolet divergent term of the Green function is energy
independent. This means that $c_r^{\MS}$ is energy independent and its
perturbative expansion is short distance dominated and can be computed
order by order in $\als$.

Note that $c_r^{\MS}$ does not depend on the long distance behavior of
$V_s^{(0)}$, only on its short distance behavior, which is universal
and dictated by perturbation theory, i.e. by
\Eqn{eq:StaticPotential}. In particular, the result is independent of
the pole mass renormalon. Therefore, the value obtained for
$c_r^{\MS}$ holds true for a general potential (not unbounded from
below at long distances) that has the correct short distance limit.

Considering the lowest order approximation of the static potential,
the Coulomb potential
\begin{equation}
V_s^{(0)}\simeq V_C=-C_F\frac{\als(\mu)}{r}
\,.
\end{equation} 
the exact solution for this potential, the Coulomb Green function, is
known in dimensional regularization and can be expressed in terms of
$\lambda\equiv C_F\, \alpha_s/\sqrt{-2 E/m_r}$ as
\begin{equation}
G_c^{(D)}(E)=\frac{g^2\, C_F\,  m^2_r}{4\pi^2}
\left(\frac{-8m_r E}{4\pi e^{-\gamma_E}}\right)^{-2\epsilon} 
\bigg[
\frac{1}{4 \epsilon}
-\frac{1}{2\lambda}+\frac{1}{2}-\gamma_E-\psi(1-\lambda)  
 +{\cal O}(\epsilon) \bigg]
\label{GCbare}
\,.
\end{equation}
According to \Eqn{eq:GD} we thus get
\begin{equation}
\label{eq:BMSC}
B^{\MS}_{V_C} = 2m_r\,C_F\,\alpha_s
\left(
-\frac{1}{2\lambda}
- \frac{1}{2} \ln\left(\frac{-8 m_r E}{\mu^2}\right)
       + \frac{1}{2} - \gamma_E - \psi(1-\lambda)
	   \right)
\,.
\end{equation}
Turning to finite-$r$ regularization the Coulomb Green function
reads
\begin{eqnarray}
G_c^{(r)}(E)
&=&
\frac{m_r^2 C_F\,\alpha_s}{\pi}
\bigg[
\frac{1}{2\, m_r C_F\,\alpha_s r_0}
-\ln \left(\mu e^{\gamma_E} r_0\right)
\nonumber
\\
&&\hspace{1cm}
-\frac{1}{2\lambda} - \frac{1}{2} \ln\left(\frac{-8 m_r E}{\mu^2}\right)
       + 1
- \gamma_E - \psi(1-\lambda)
\bigg]\,.
\label{eq:rGreenFunc1}
\end{eqnarray}
whereas $u_0(r_0)$ for the Coulomb case is given by
\begin{equation}
u_0(r_0) = 1 - 2m_r\,r_0\, C_F\,\alpha_s\,
 \ln\left(\mu e^{\gamma_E}r_0\right)\, .
\end{equation}
The expression stops at ${\cal O}(\als)$. 
In the Coulomb approximation there are no ${\cal O}(\als^2)$ terms in
$u_0(r_0)$ and, therefore, in $A_{V_C}^{(r)}$.  Using \Eqn{eq:Gr} we then find
\begin{equation}
\label{eq:BRC}
B_{V_C}^{(r)}(E) = 
2 m_r\,C_F\,\alpha_s
\left[
-\frac{1}{2\lambda}
- \frac{1}{2} \ln\left(\frac{-8 m_r E}{\mu^2}\right)
       + 1 - \gamma_E - \psi(1-\lambda)
	   \right]
\,.
\end{equation}
Note that in an strict NNLO or NNLL computation of the decay ratio
this would be the only term that should be considered.

Thus we compute $c^{\MS}_r$ in an expansion in $\alpha_s$  and
obtain\footnote{If the constant $e^{\gamma_E}$ were not introduced in
  Eq. (\ref{eq:Ar}), $c^{\MS}_r$ would read
\begin{equation}
\label{eq:CDbis}
c^{\MS}_{(r)}=-2m_rC_F\als\left(\frac{1}{2}-\gamma_E\right)
+{\cal O}(\als^2)
\,.
\end{equation}}
\begin{equation}
c^{\MS}_r=-2m_r\frac{C_F\alpha_s}{2}+{\cal O}(\alpha_s^2).
\label{eq:CD}
\end{equation}
This constant can also be obtained from the difference between
dimensional- and $r$-regularized computations at finite order in
$\als$. At the lowest order it corresponds to the computation of one-
and two-loop contributions to the Green function in both schemes, by
considering the difference of their renormalized pieces. We have
checked in an explicit calculation that \Eqn{eq:CD} is reproduced by
the difference of two-loop contributions.

In order to obtain the ${\cal O}(\als^2)$ corrections to
$c^{\MS}_{r}$ one has to include the ${\cal O}(\als^2)$ corrections
to the static potential and compute the associated corrections to the
Green function in both schemes. In principle, this is possible and
partial results can be found in the literature. Nevertheless, this
would go beyond the aim of this work, since it would produce
corrections that are anyway unmatched by the precision of the hard
matching coefficient.

Finally, for a general potential with the right short distance
structure, we can combine \Eqn{eq:rho_formulaMS} with
\Eqns{eq:BD_Br}{eq:CD} and write
\begin{equation}
\label{deltaMSKiyo}
\delta \rho_n^{\MS}(\mu) =
-\frac{8 m_r C_F}{3 m_1m_2}
D_{S^2,s}^{(2)}(\mu)
\left(\widehat{B}_{V^{(0)}_s}^{(r)}(E^{(0)}_n;\mu) +
\frac{1}{3} m_rC_F\,\als+{\cal O}(\als^2)\right).
\end{equation}
Once we know the $\MS$ expression we can also write $\delta \rho_n$ 
in different schemes. For instance, in the "hard-matching" scheme used 
in Ref. \cite{Penin:2004ay} we have
 \begin{equation}
\label{deltaHMKiyo}
\delta \rho_n^{HM}(\mu) =
-\frac{8 m_r C_F}{3 m_1m_2}
D_{S^2,s}^{(2)}(\mu)
\left(\widehat{B}_{V^{(0)}_s}^{(r)}(E^{(0)}_n;\mu) - 
2m_rC_F\,\als+{\cal O}(\als^2)\right),
\end{equation}
which will be relevant afterwards.

These results enable us to compute the decay ratio in terms of
$\widehat{B}_{V^{(0)}_s}^{(r)}$, whose determination will be discussed
in the next section.

\section{Determination of $\widehat{B}^{(r)}_{V_s^{(0)}}$}
\label{sec:BrVs}

In this section we determine
$\widehat{B}^{(r)}_{V_s^{(0)}}(E^{(0)}_n)$ in several approximation
schemes for $V_s^{(0)}$.  We have already mentioned that our idea is
to treat the static potential exactly, yet we only know its expression
up to three loops. There is some freedom on how this truncation is
performed. This produces a class of potentials to study, which
introduces some scheme and scale uncertainties.  As we have stressed
in previous sections, the analysis applies to any arbitrary potential
with the correct short distance behavior and not unbounded from below
at long distances.  Therefore, in what follows we will consider
different approximations to the static potential. One quality that
they have in common is the renormalon cancellation. We have to
preserve renormalon cancellation between the static potential and the
pole mass of the heavy quark. At the same time we will be forced to
consider the resummation of logarithms to reproduce the correct
behavior of the potential at short distances. Thus we will have to
devise schemes where both the resummation of the logarithms and the
renormalon cancellation is achieved order by order in the perturbative
expansion. We illustrate this discussion in the following sections,
where we show the determination of $B^{(r)}_{V_s^{(0)}}$ using either
the Coulomb potential, the static potential at different orders in
$\als(\mu)$, and the static potential at different orders in
$\als(1/r)$. In this last case we will use different schemes with
renormalon cancellation.  The dependence on the scheme of renormalon
subtraction (potential) may give an estimate of the error, since it is
also a measure of the dependence on the long distance behavior of the
potential.

Finally, let us note as well that, in order for our computation to
make sense, the successive approximations to the static potential
should be convergent (or at least small) themselves. We will check
this convergence in this section.

\subsection{Coulomb potential}

If we approximate the static potential by the Coulomb potential $V_C$
we can get an analytic solution for $\widehat{B}^{\MS}_{V_C}$ by
directly working in dimensional regularization.  Expanding
$G_c^{(D)}(E)$ as given in \Eqn{GCbare} around its poles at $E^{(0)}_n
\equiv - m_r C_F \als^2/(2n^2)$ we can write
\begin{equation}
\label{Greenorig}
G^{(D)}_c(E)
= -\frac{\alpha_s\, C_F\, m_r^2}{\pi}
\frac{2\, E^{(0)}_n}{n(E^{(0)}_n-E)}
+\widehat G_c^{(D)}(E^{(0)}_n) + {\cal O}(E-E^{(0)}_n)
\end{equation}
with
\begin{equation}
\widehat G_c^{(D)}(E^{(0)}_n) =
\frac{g^2\, C_F\, m_r^2}{4\pi^2}
\left(\frac{-8m_rE}{4\pi e^{-\gamma_E}}\right)^{-2\epsilon}
\bigg[
\frac{1}{4\epsilon}
+ \frac{1}{2} - \gamma_E + \frac{1}{n}-\psi(n)
+{\cal O}(\epsilon) \bigg]\, .
\end{equation}
Comparing to \Eqn{eq:GDhat} we obtain
\begin{equation}
\label{BdCoulomb}
\widehat B_{V_C}^{\MS}(E^{(0)}_n)
=
2m_rC_F\als
\left(
- \frac{1}{2} \ln \frac{-8 m_r E^{(0)}_n}{\mu^2}
       + \frac{1}{2} - \gamma_E + \frac{1}{n}-\psi(n)
	   \right)
\end{equation}
and thus $\delta \rho_n^{\MS}(\mu)$ in the Coulomb approximation
directly from \Eqn{eq:rho_formulaMS}.

\subsection{Fixed order $V_s^{(0)}$: $\alpha_s(\mu_s)$ expansion}
\label{sec:alsmuexp}

\begin{figure}[ht]
\begin{center}
\epsfxsize=13cm
\epsffile{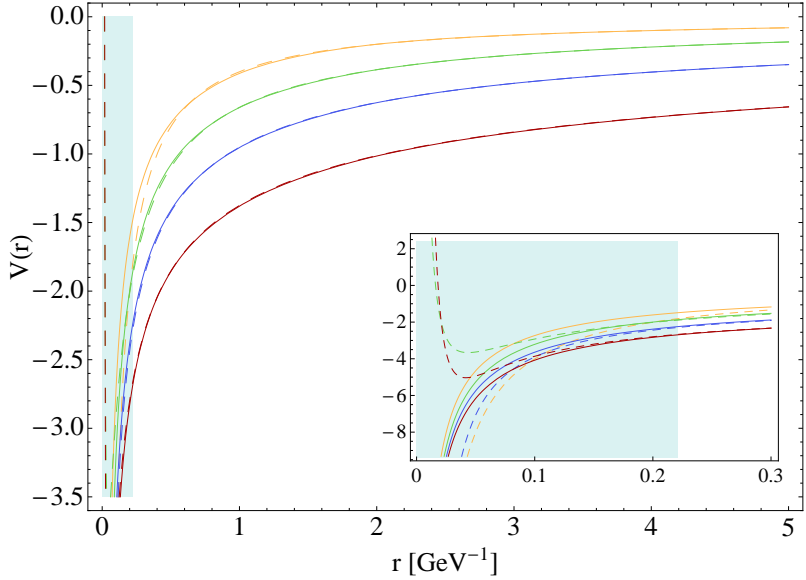}
\end{center}
\caption{The FO (dashed) and RGI (solid) static potential $V_{SD}(r)$ 
according to \Eqn{VSDfo} and \Eqn{VSDrg} respectively. We take $\mu_s =
  2$~GeV, $N_l=4$ and $\mu_r=2$~GeV. The potential is shown as a
  function of $r$ at LO (yellow), NLO (green), NNLO (blue) and NNNLO
  (red) with the small $r$ region shown in the inset. The shaded area
  in blue indicates the short distance regime $0<r<1/m_b$.}
\label{fig:potential}
\end{figure}

The standard way to go beyond the Coulomb potential approximation for
the static potential is to make an expansion in $\als(\mu_s)$. Thus we
write
\begin{equation}
\widetilde{V}_{SD}(q) =
-\frac{4\pi C_F\, \alpha_s(\mu_s)}{{\bf q}^2}
\,
\bigg(
1+ \sum_{n=1}^{\infty}\bigg(\frac{\alpha_s(\mu_s)}{4\pi}\bigg)^n\, 
\widetilde{a}_n(\mu_s;q)\bigg).
\label{eq:VtildeSDfo}
\end{equation}
This expanded version of the static potential is often used in
quarkonium phenomenology to respect rigorous expansion according to
non-relativistic power counting\footnote{In the most rigorous fixed
  order computation only the Coulomb part of the static potential is
  treated exactly and $\alpha_s$ corrections corresponding to the
  second and remaining terms in Eq.(\ref{eq:VtildeSDfo}) are treated
  iteratively order by order by insertion.}.
In position space we have
\begin{eqnarray}
\lim_{r \rightarrow 0}V_s^{(0)}(r)=V_{SD}(r)
&=&
 -\frac{C_F\,\alpha_s(\mu_s)}{r}\,
\bigg\{1+\sum_{n=1}^{\infty}
\left(\frac{\alpha_s(\mu_s)}{4\pi}\right)^n a_n(\mu_s;r)\bigg\}
\label{VSDfo}
\end{eqnarray}
In practice we will take the static potential up to NNNLO, i.e. up to
${\cal O}(\als^4)$ including also the leading ultrasoft corrections.
This means we take into account the first three terms of this
expansion with coefficients
\begin{eqnarray}
a_1(\mu_s;r)
&=&
a_1+2\beta_0\,\ln\left(\mu_s e^{\gamma_E} r\right)
\,,
\nonumber\\
a_2(\mu_s;r)
&=&
a_2 + \frac{\pi^2}{3}\beta_0^{\,2}
+\left(\,4a_1\beta_0+2\beta_1 \right)\,\ln\left(\mu_s e^{\gamma_E} r\right)\,
+4\beta_0^{\,2}\,\ln^2\left(\mu_s e^{\gamma_E} r\right)\,
\,,
\nonumber \\
a_3(\mu_s;r)
&=&
a_3+ a_1\beta_0^{\,2} \pi^2+
\frac{5\pi^2}{6}\beta_0\beta_1 +16\zeta_3\beta_0^{\,3}
\nonumber \\
&+&\bigg(2\pi^2\beta_0^{\,3} 
+ 6a_2\beta_0+4a_1\beta_1+2\beta_2
+\frac{16}{3}C_A^{\,3}\pi^2\bigg)\,
  \ln\left(\mu_s e^{\gamma_E} r\right)\,
\nonumber \\
&+&\bigg(12a_1\beta_0^{\,2}+10\beta_0\beta_1\bigg)\,
  \ln^2\left(\mu_s e^{\gamma_E} r\right)\,
+8\beta_0^{\,3}  \ln^3\left(\mu_s e^{\gamma_E} r\right)\,
\nn
\\
&+&\delta a_3^{us}(\mu_s,\mu_{us}),
\label{eq:Vr}
\end{eqnarray}
Explicit expression for $a_i(\mu_s;r)$ can be found in the literature
\cite{FSP,Schroder, short,KP1,RG,Anzai:2009tm,Smirnov:2009fh}. For the
ultrasoft corrections to the static potential we take
\begin{equation}
\delta a_3^{us}(\mu_s,\mu_{us})
\simeq \frac{16}{3}C_A^3 \pi^2\ln\left(\frac{\mu_{us}}{\mu_s}\right)
\, .
\end{equation}
We will not consider the renormalization group improved ultrasoft
contribution in this paper as its numerical impact is small.  The
potential is shown in Figure~\ref{fig:potential} (dashed lines) for
$\mu_s = 2$~GeV and the number of light flavors set to $N_l=4$.  It is
clear that for small $r$, depicted in the inset of
Figure~\ref{fig:potential}, there are serious issues regarding the
convergence.  The potential changes drastically in going from LO to
NLO to NNLO etc. This behavior occurs for the typical values of
$\mu_s$ and $N_l$ that apply for the charm and bottom case. As one
increases the value of $\mu_s$, one has to go to shorter distance to
see this effect, as it would happen for top.

Ignoring this problem for the moment and working with the Fixed Order
(FO) static potential we can obtain $u_0(r_0)$ and, therefore,
$A^{(r)}_{V_s^{(0)}}(r_0)$ as an expansion in $\als$ as well. We find
\begin{eqnarray}
A^{(r)}_{V_s^{(0)}}(r_0)
&=&
\frac{1}{r_0}
-2m_r \, \alpha_s(\mu)\,C_F\, v(l_0)
\,,
\nonumber
\\
v(l_0)
&=&
\sum_{i=0}^{3} v_n\left(l_0\right)\,
         \left(\frac{\alpha_s(\mu)}{4\pi}\right)^n
         \,,
\label{eq:rSingularGreenFunction}
\end{eqnarray}
where $l_0=\ln\left(\mu\, e^{\gamma_E}r_0\right)$ and the expansion
coefficients are given by
\begin{eqnarray}
v_0(l_0)&=&l_0
\,,
\nonumber\\
v_1(l_0)
&=&
 \big(a_1-2\beta_0\big)\,l_0
+\beta_0\,l_0^{\,2}
\,,
\nonumber\\
v_2(l_0)
&=&
\bigg(a_2-4a_1\beta_0+8\beta_0^{\,2}
+\frac{\pi^2}{3}\beta_0^{\,2}-2\beta_1\bigg)\, l_0
\nonumber \\
&+& \bigg(2a_1\beta_0-4\beta_0^{\,2}+\beta_1\bigg)\, l_0^{\,2}
+\frac{4\beta_0^2}{3}\,l_0^3
\,,
\nonumber \\
v_3(l_0)
&=&
\bigg(a_3 + \delta a_3^{us}
      -6a_2\beta_0+24a_1\beta_0^{\,2}+a_1\beta_0^{\,2}\pi^2
      -\left(48+2\pi^2\right)\beta_0^{\,3}-4a_1\beta_1
\nonumber \\
&&     +\left(20+\frac{5\pi^2}{6}\right)\beta_0\beta_1
      -2\beta_2+16\beta_0^{\,3}\zeta_3
       -\frac{16}{3}\pi^2C_A^{\,3}
\bigg)\,l_0
\nonumber \\
&+&
\bigg(3a_2\beta_0-12a_1\beta_0^{\,2}+\left(24+\pi^2\right)\beta_0^{\,3}
       +2a_1\beta_1-10\beta_0\beta_1+\beta_2 +\frac{8\pi^2}{3}C_A^{\,3}
\bigg)\,l_0^{\,2}
\nonumber\\
&+&
\bigg(4a_1\beta_0^{\,2}-8\beta_0^{\,3}+\frac{10}{3}\beta_0\beta_1
\bigg)\, l_0^{\,3}
+2\beta_0^{\,4}\,l_0^{\,4}.
\end{eqnarray}
The $\mu$ dependence appearing in \Eqn{eq:rSingularGreenFunction}
enters through \Eqn{eq:C0} and should be cancelled in \Eqn{Rdef}. Even
though the exact expression for the static potential is scale
independent, working at a finite order in $\als(\mu_s)$ there is some
residual $\mu_s$ dependence.

The computation of $B^{(r)}_{V_s^{(0)}}$ is done numerically along the
lines of Section~\ref{Gr}. We use the input values $m_{b,\rm
  PS}(2\,{\rm GeV})=4.515\,{\rm GeV}$~\cite{Pineda:2006gx} and
$m_{c,\rm PS}(0.7\,{\rm GeV})=1.50\,{\rm GeV}$~\cite{Signer:2008da}
for bottom and charm quarks, respectively.  They can be translated
into scale-invariant $\MS$-mass of $\overline{m}_b=4.19\,{\rm GeV}$
and $\overline{m}_c=1.25\,{\rm GeV}$.  The strong coupling
$\alpha_s^{(n_f=5)}(M_z)=0.118$ is used as an input evolved down to
low energy scale using 4-loop running formulae.  For the top quark
mass we use $m_{t, \rm PS}(20\,{\rm GeV}) = 173$~GeV for
illustration. The scale $\mu_{us}$ needed for the leading ultrasoft
contribution is set to $\mu_{us}=0.7$~GeV for charm, $\mu_{us}=1$~GeV
for bottom and $\mu_{us}=10$~GeV for top.

\begin{figure}
\epsfxsize=10cm
\centerline{\epsffile{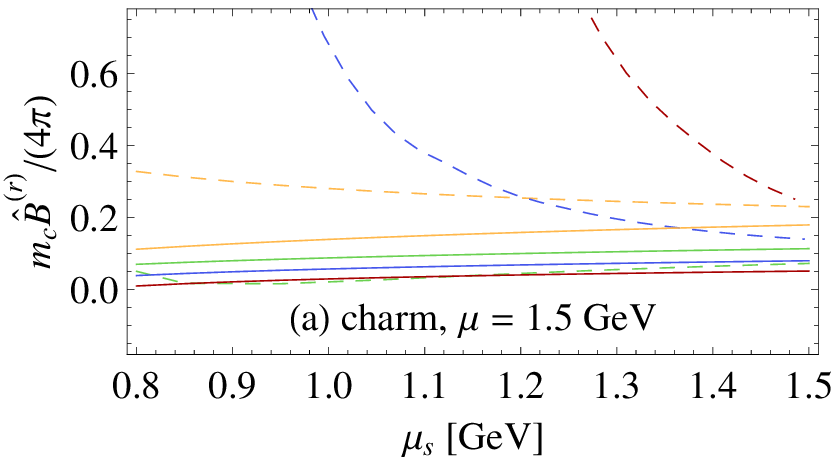}}
\medskip
\epsfxsize10cm
\centerline{\epsffile{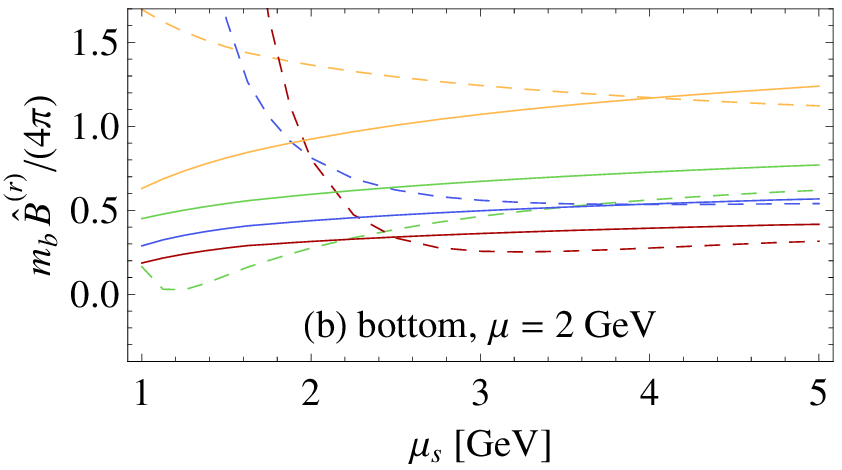}}
\medskip
\epsfxsize=10cm
\centerline{\epsffile{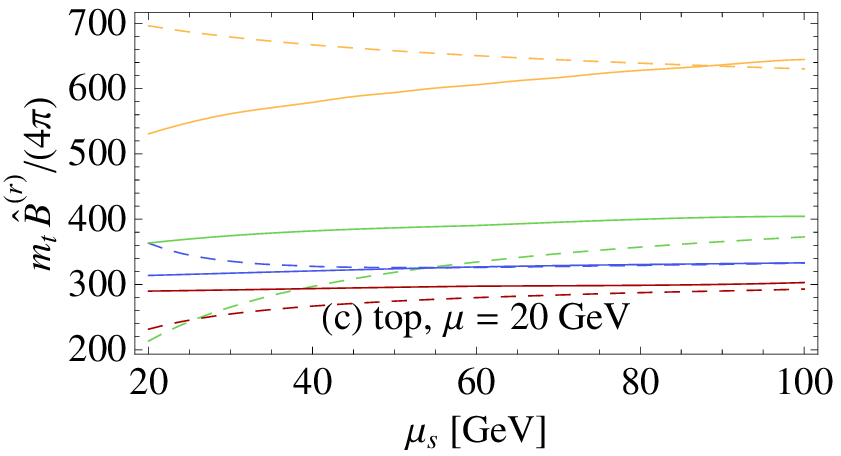}}
\caption{$\widehat{B}^{(r)}_{V_s^{(0)}}$ as
  a function of $\mu_s$ at LO(yellow), NLO(green), NNLO(blue) and
  NNNLO(red) with $\mu = 1.5$ GeV for charm, 2 GeV for bottom, and 
  20 GeV for top. Dashed lines are obtained using
  \Eqn{VSDfo}, solid lines are obtained using \Eqn{VSDrg}. }
\label{fig:Brfixed}
\end{figure}

In Figure~\ref{fig:Brfixed} we show the results for charm, bottom and
top (dashed lines) as a function of the scale $\mu_s$ for fixed
$\mu$. For illustration we have chosen $\mu = 1.5$~GeV for charm,
2~GeV for bottom and 20~GeV for top.  Note that, ideally, the result
should be independent of $\mu_s$, as it reflects a dependence on the
long distance behavior of the potential.  For charm and bottom we see
problems of convergence, in particular for small values of
$\mu_s$. This is due to the behavior of the potential at short
distances, which we have already illustrated in
Figure~\ref{fig:potential}. For top the situation is much better. Note
that the LO curve corresponds to the Coulomb potential. In all three
cases we observe a significant gap between the Coulomb solution and
the higher order corrections (for the range of $\mu_s$ for which the
result can be trusted).

Before we address the problem of the bad convergence, let us remark
that expanding the potential in $\als(\mu)$, the pole mass renormalon
enters as an $r$-independent constant in the potential. This constant
cancels in the evaluation of $B^{(r)}_{V_s^{(0)}}$, which is
independent of the overall normalization of the potential. Thus, in
this evaluation the dependence will only enter in the values of the
mass used. The error associated to this uncertainty is beyond our
accuracy.

\subsection{RG-Improved $V_s^{(0)}$: $\als(1/r)$ expansion}
\label{sec:RGIpot}

In the previous subsection we have seen that the convergence for
$\widehat{B}^{(r)}_{V_s^{(0)}}$ is very unsatisfactory if we use
\Eqn{VSDfo}. Surprisingly the problem comes from short and not long
distances.  The solution is to absorb the large logarithms into the
running coupling.  However, this has to be done carefully in order not
to destroy the renormalon cancellation achieved order by order in
$\als$. More specifically, we consider different approximations to the
static potential behaving for $r \rightarrow 0$ as
\begin{equation}
V_s^{(0)} \simeq -\frac{C_F\,\alpha_s(1/r)}{r}
\bigg\{ 1+ \sum_{n=1}^{3}a_n(1/r;r)
   \left(\frac{\als(1/r)}{4\pi}\right)^n
\bigg\}
\,,
\label{eq:Vsresum}
\end{equation}
and yet achieving renormalon cancellation order by order in
$\als^n(1/r)$. This will give us an estimate of the dependence of the
result on the long distance behavior of the potential. We will
generically name this class of potentials renormalization group
improved (RGI) and denote them by LO, NLO, ...  according to the power
of $\als(1/r)$ at which we stop the perturbative expansion in
\Eqn{eq:Vsresum}.

One possibility that fulfills all these requirements is the PS scheme
\cite{Beneke:1998rk} with the following modification:
\begin{equation}
\label{VPS}
V_{PS}(r) = V_{SD}(r,\mu_r)+2\,\delta m_{PS}
\end{equation}
with
\begin{equation}
V_{SD}(r,\mu_r) \equiv 
\int_{q\leq \mu_r} \frac{d^3 {\bf q}}{(2\pi)^3}
  e^{i{\bf q}\cdot{\bf r}}
  \widetilde{V}_{SD}|_{\mu=\mu_s}(q)
+
\int_{q > \mu_r} \frac{d^3 {\bf q}}{(2\pi)^3}
  e^{i{\bf q}\cdot{\bf r}}
 \widetilde{V}_{SD}|_{\mu=q}(q)
\label{VSDrg}
\end{equation}
Thus we introduce a factorization scale $\mu_r$. For $q<\mu_r$ we
expand $\widetilde{V}_{SD}(q)$ in $\als(\mu_s)$, as in the previous
subsection. For $q>\mu_r$ however, we use the running coupling in
$\widetilde{V}_{SD}(q)$. As can be seen in Figure~\ref{fig:potential},
the RGI potential (solid lines) shows good convergence for all values
of $r$. We have checked that the results for
$\widehat{B}^{(r)}_{V_s^{(0)}}$ are not sensitive to the precise value
of the factorization scale, as long as $\mu_r$ is large enough.  This
definition has the advantage that the renormalon contribution is $r$
independent and achieves the resummation of logarithms. The fact that
the renormalon cancellation is $r$ independent makes it possible to
work also with $\delta m_{PS}=0$ in \Eqn{VPS}, as far as the
determination of $\widehat{B}^{(r)}_{V_s^{(0)}}(E^{(0)}_n)$ is
concerned.

The results for $\widehat{B}^{(r)}_{V_s^{(0)}}$, using \Eqn{VSDrg}
rather than \Eqn{VSDfo} are depicted as solid lines in
Figure~\ref{fig:Brfixed}. We have taken $\mu_r = 1$~GeV, $\mu =
1.5$~GeV for charm, $\mu_r = \mu= 2$~GeV for bottom, and $\mu_r =\mu=
20$~GeV for top. As can be seen, the resummation of logarithms results
in a dramatic improvement in the charm and bottom case, and is also
quite significant in the top case. In all three cases, the RG result
is nearly independent of $\mu_s$.  This signals a weak dependence on
the long distance tail of the potential.  This is to be contrasted
with the results obtained using \Eqn{VSDfo}, which are completely
unreliable unless unnaturally large values for $\mu_s$ are used
(especially for charm).  The RGI curves show a good convergent pattern
for top, and also a reasonable convergence in the case of bottom.
Even for charm we see signs of convergence, albeit marginal.  In
particular, in this case, and to a lesser extent in the case of
bottom, the splitting between the NNLO and NNNLO curves is not much
smaller than the splitting between the NLO and NNLO curves. Note,
though, that at NNNLO the potential starts to be sensitive to
ultrasoft physics, which we do not include in our analysis. In this
respect the NNNLO curves are to be considered incomplete (though the
explicit dependence on the ultrasoft factorization scale is
small). Moreover, at some point the asymptotic behavior of the
perturbative series should set in and it cannot be ruled out that we
are approaching this regime. Still, we would like to point out the
smallness of this splitting compared to the total magnitude of the
correction achieved by the reorganization of the perturbative series,
which can be estimated by comparing the Coulomb line versus the NNNLO
curve. In this respect, even if we consider the splitting between the
NNLO and NNNLO curves as an error, its magnitude is rather small
compared with the total gap. From this analysis, we conclude that we
should use the RGI potential instead of the FO one and we will take
this attitude in the rest of the paper.

Another possibility that we explore is the use of the RS or RS'
potential \cite{Pineda:2001zq}.  To avoid numerical instabilities, due
to the behavior of the potential at long distances, we also modify the
potential in the following way:
\begin{equation}
\label{VRS}
V_{\RS}(r)=
\,\left\{
\begin{array}{ll}
&
\displaystyle{
(V_{SD}+2\delta m_{\RS})|_{\mu=\mu_s}=
 \sum_{n=0}^{\infty}V_{RS,n}\als^{n+1}(\mu_s)
\qquad {\rm if} \quad r>\mu_r }
\\
&
\displaystyle{
(V_{SD}+2\delta m_{\RS})|_{\mu=1/r}=
 \sum_{n=0}^{\infty}V_{RS,n}\als^{n+1}(1/r)
\qquad {\rm if} \quad r<\mu_r }
\end{array} \right.
\end{equation}
Irrespectively of the potential we use, the short distance behavior of
the potential and, consequently, $A^{(r)}(r_0;\mu)$ is the same. The
full expression for $A^{(r)}(r_0;\mu)$ is more complicated in these
cases than in Section~\ref{sec:alsmuexp} and we refrain 
from giving the general explicit expression
and only show (for illustration) how it would look like
at the lowest order. If for instance we consider LL
running at short distance, namely
\begin{eqnarray}
V_s^{(0)}(r)\simeq 
-\frac{C_F}{r}\,
  \frac{\alpha_s(\mu)
        }{1-\frac{\beta_0\alpha_s}{2\pi} \ln(\mu r)}
\,,
\end{eqnarray}
we have
\begin{eqnarray}
A^{(r)}(r_0)
&=&
\frac{1}{r_0}
-2m_r\,C_F\,\alpha_s(\mu)\, v(l_0)
\,.
\label{ArRG}
\end{eqnarray}
\begin{eqnarray}
v(l_0)
&=&
\frac{2\pi}{\beta_0\alpha_s(\mu)}
\bigg\{
f\bigg [\gamma_E+\frac{2\pi}{\beta_0\alpha_s(\mu)}-l_0\bigg ]
-f\bigg [\gamma_E+\frac{2\pi}{\beta_0\alpha_s(\mu)}\bigg ]
\bigg\}
\,,
\end{eqnarray}
with
$f[x ]\equiv e^x\,{\rm Ei}(-x)-\ln\,x$.
The coefficients $v_n$ have an expansion in $\alpha_s$ and $l_0$
\begin{eqnarray}
v(l_0)
&=&
l_0
+\left(\frac{\beta_0\alpha_s}{4\pi}\right)\,
 \bigg\{-2(\gamma_E+1)l_0+l_0^{\,2}\bigg\}
\nonumber
\\
&+&\left(\frac{\beta_0\alpha_s}{4\pi}\right)^2\,
 \bigg\{\left(8+8\gamma_E+4\gamma_E^2\right)\,l_0
       -(4+4\gamma_E)l_0^{\,2}+\frac{4}{3}l_0^{\,3}
 \bigg\}+\cdots
\,.
\end{eqnarray}

\begin{figure}
\epsfxsize=10cm
\centerline{\epsffile{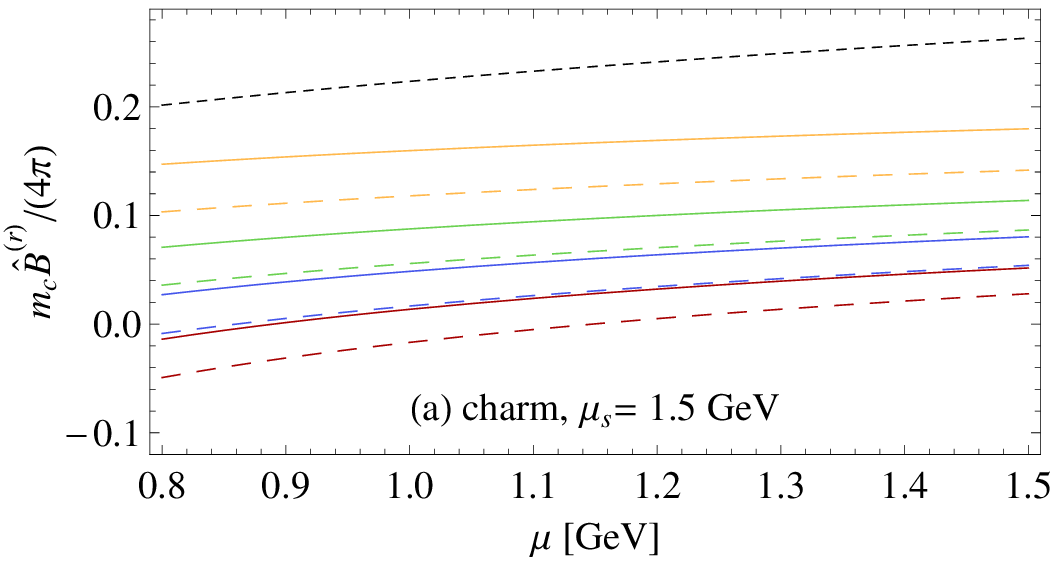}}
\medskip
\epsfxsize10cm
\centerline{\epsffile{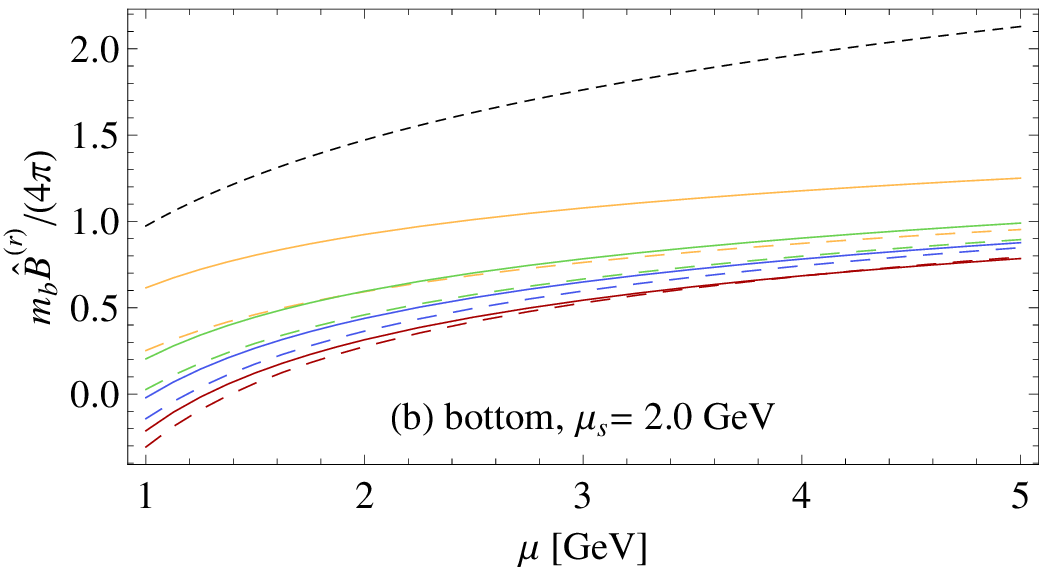}}
\medskip
\epsfxsize=10cm
\centerline{\epsffile{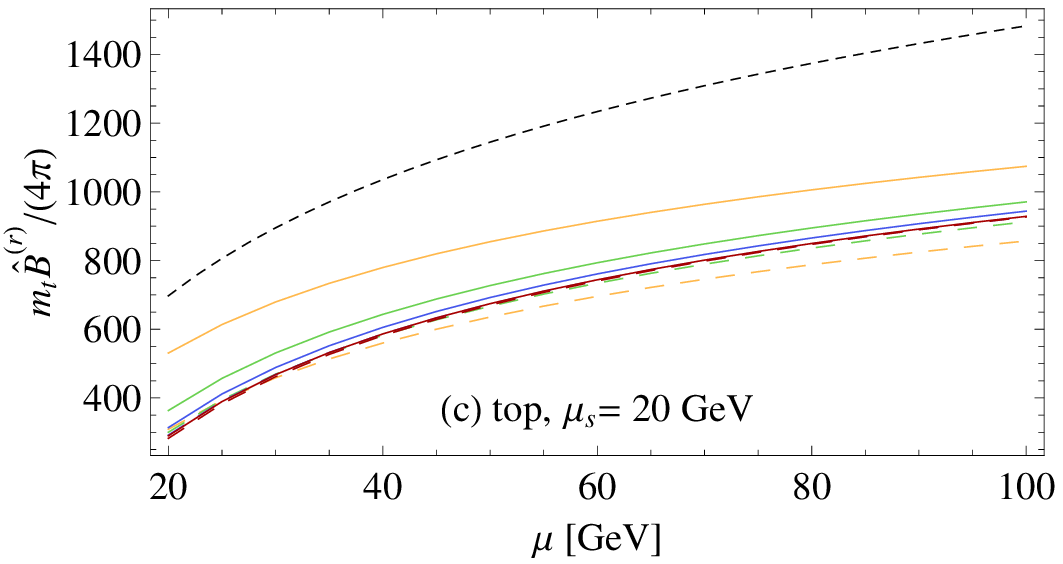}}
\caption{\label{fig:Brscheme2_b} $\widehat{B}^{(r)}_{V_s^{(0)}}$ as a
  function of $\mu$ at LO (yellow), NLO (green), NNLO (blue) and NNNLO
  (red) with $\mu_s = 1.5$ GeV, $\mu_r = 1$~GeV and
  $\mu_F=\mu_{us}=0.7$~GeV for charm, $\mu_s=\mu_r =\mu_F= 2$~GeV and
  $\mu_{us}=1$~GeV for bottom, and $\mu_s=\mu_r = \mu_F=20$~GeV, and
  $\mu_{us}=10$~GeV for top. Solid lines are obtained in the PS scheme
  using \Eqn{VSDrg} and dashed lines are obtained in the RS' scheme
  using \Eqn{VRS}. For reference we also include
  $\widehat{B}^{(r)}_{V_C}$ (short-dashed black line).}
\end{figure}

We now perform the numerical evaluation of
$\widehat{B}^{(r)}_{V_s^{(0)}}$ at different orders in the static
potential and compare the results obtained using the PS and RS'
scheme. The results are shown in Figure~\ref{fig:Brscheme2_b}. The
difference between the schemes is small and converging for the case of
bottom and top.  In these two cases the differences between both
schemes is pretty small for the NNNLO curves. This is again a good
signal, since the dependence on the scheme is an indirect measure of
the dependence on the long distance tail of the potential. For charm,
the situation is less convincing. The gaps between schemes show
marginal convergence at best as we increase the order. Yet, this gap
is still much smaller than the gap between the Coulomb result and the
NNNLO result. Comparing the Coulomb result, shown as the black short-dashed
line, to our results, we can see that in all three cases a rather
significant portion of the correction is already achieved with the LO
RGI potential. In the case of top the NLO RGI potential is already
quite close the most accurate NNNLO result. This behavior is also
seen, to a lesser extent in the case of bottom.  Note that the LO RGI
potential exactly incorporates the $r$-dependent leading logarithms.
This is equivalent to introducing an infinite number of corrections to
the Coulomb potential and to iterate them an infinite numbers of
times. This reorganization of perturbation theory seems to produce a
major effect.  Another observation is that the RS' scheme produces an
accelerated convergence to the asymptotic regime.  This is clearly
seen in the top case, and to a lesser extent, in the bottom case. In
those cases the LO RGI potential produces the bulk of the correction and
the magnitude of the higher order corrections is smaller in the RS'
than in the PS scheme. The price paid is that the splitting between
different orders in the RS' scheme is less convergent.

The dependence of the results on $\mu_r$ is very small. Changing
$\mu_r$ from 2~GeV to 4~GeV for example results in differences that
are an order of magnitude smaller than the changes we find by going
from say LO to NLO.  

The dependence on $\mu$ will have to be cancelled by the scale
dependence of the matching coefficient $c_s(\mu)$.  Note that our
evaluation of $\widehat{B}^{(r)}_{V_s^{(0)}}$ also includes subleading
logarithms, which are not matched by the precision of the RG (hard)
computation. The fact that the scale dependence roughly corresponds to
the Coulomb potential (with RG running) can be taken as an indication
that subleading logarithms are not very important (see
Figure~\ref{fig:Brscheme2_b} for illustration).

\begin{figure}
\epsfxsize=10cm
\centerline{\epsffile{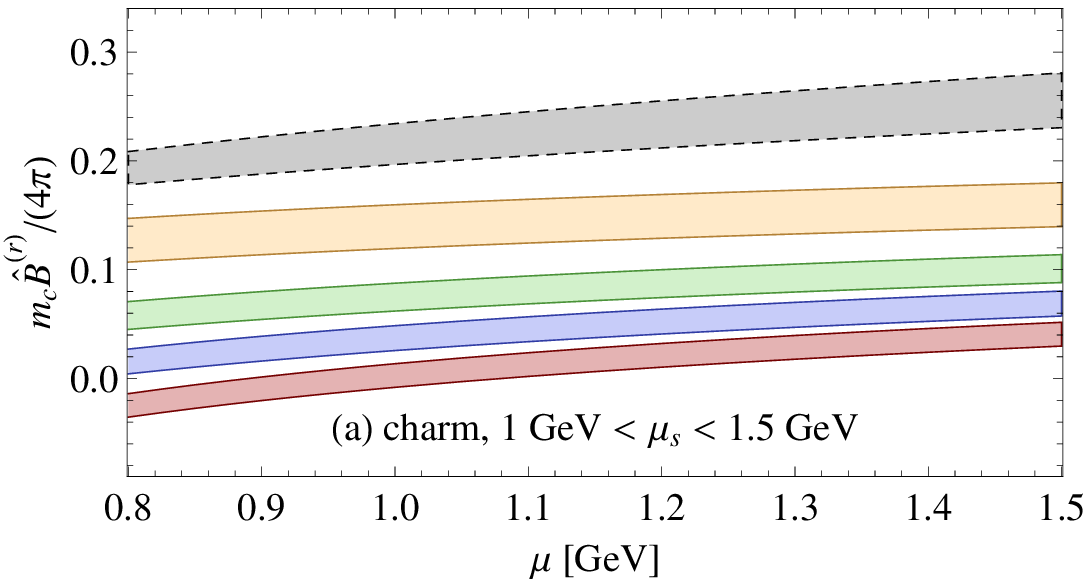}}
\medskip
\epsfxsize10cm
\centerline{\epsffile{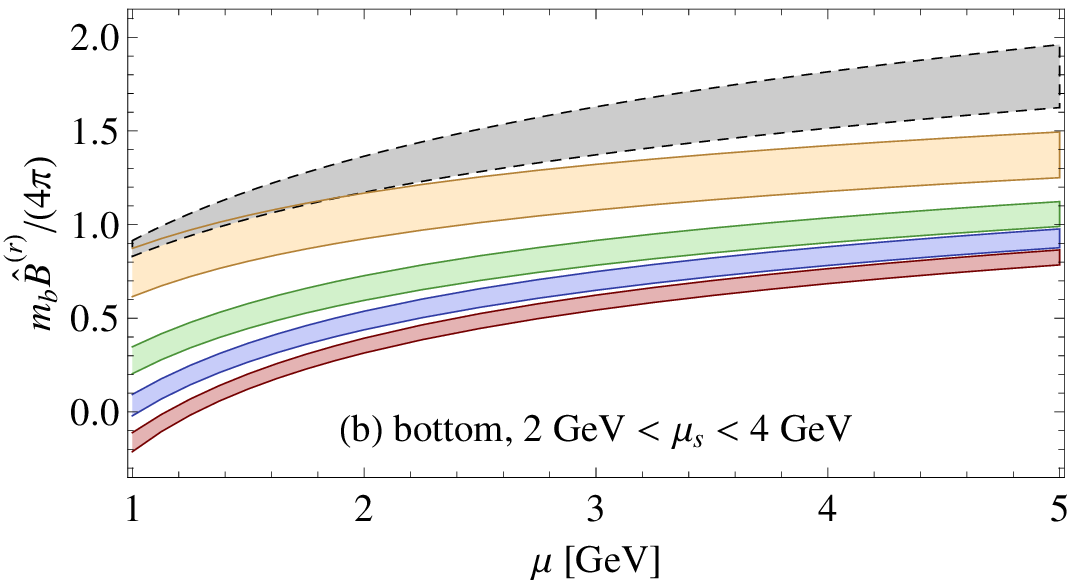}}
\medskip
\epsfxsize=10cm
\centerline{\epsffile{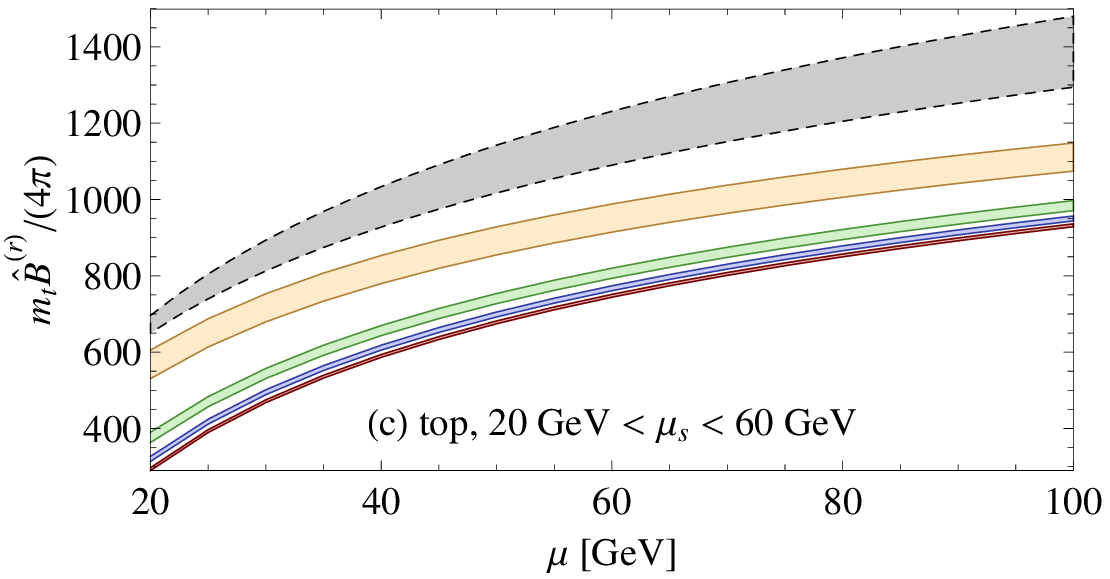}}
\caption{\label{fig:Brmus} $\widehat{B}^{(r)}_{V_s^{(0)}}$ using
  \Eqn{VSDrg} as a function of $\mu$ at LO (yellow), NLO (green), NNLO
  (blue) and NNNLO (red) with $\mu_r=1$~GeV and
  $\mu_F=\mu_{us}=0.7$~GeV for charm, $\mu_r =\mu_F= 2$~GeV and
  $\mu_{us}=1$~GeV for bottom, and $\mu_r = \mu_F=20$~GeV, and
  $\mu_{us}=10$~GeV for top. The bands are obtained by variation of
  $\mu_s$ in the range 1--1.5 GeV, 2--4 GeV and 20--60 GeV for charm,
  bottom and top respectively.  For reference we also include
  $\widehat{B}^{(r)}_{V_C}$ (grey band).}
\end{figure}

Finally, there is also a dependence on the scale $\mu_s$.  This
dependence (as the dependence on the renormalon subtraction scheme)
partly reflects the dependence of the result on the long distance tail
of the potential. On the other hand one can not take $\mu_s$ very
small otherwise $\als(\mu_s)$ becomes very large. We now perform the
numerical evaluation of $\widehat{B}^{(r)}_{V_s^{(0)}}$ at different
orders in the loop expansion in the PS scheme and using different
values of $\mu_s$. The results for varying values of $\mu_s$ are shown
as bands in Figure~\ref{fig:Brmus}. We also show the Coulomb result as
the band enclosed by black dashed lines. This plot also illustrates
that the bulk of the correction is already achieved with the LO/NLO
RGI potential in the top and bottom case, where we have convergence
(in the charm case convergence is marginal at best). The $\mu_s$
dependence tends to diminish as one increases the order of the RGI
potential in the top, and to a lesser extent in the bottom case. In
the charm case the $\mu_s$ dependence remains almost
constant. Overall, we find that the $\mu_s$ dependence is slightly
larger than the scheme dependence, but still smaller than the typical
gap due to working at different orders in the RGI potential.

\section{Phenomenology of the decay ratio}

Using the results obtained for ${\hat
  B}^{(r)}_{V_s^{(0)}}(E^{(0)}_n;\mu)$ we can get improved
determinations of the decay ratio, by combining Eqs.~(\ref{Rdef}),
(\ref{rho_n}) and (\ref{deltaMSKiyo}) with the determination of $c_s$
from Ref.~\cite{Penin:2004ay}.  We use the results obtained in
Section~\ref{sec:RGIpot} with the RGI potential, since they both
achieve the resummation of logarithms and renormalon cancellation. The
main source of uncertainties in the evaluation of ${\hat
  B}^{(r)}_{V_s^{(0)}}(E^{(0)}_n;\mu)$ is reflected by the
computations at different orders in $\als$ in the static potential
and, to a lesser extent, by the dependence on $\mu_s$.  In comparison,
the dependence on the quark mass, $\mu_r$, $\mu_f$ and $\mu_{us}$ is
small. Therefore, we will fix those parameters to the values used in
Section~\ref{sec:RGIpot}.  In Section~\ref{sec:RGIpot} we also saw
that the scheme dependence for renormalon subtraction was small,
compared with the uncertainty due to the computation at different
orders.  Therefore, we will only take one scheme (PS) for reference in
the plots.

In order to explore different power counting expansions for our
results, we will consider and compare different approximations.  In
particular we will show the effect of resumming logarithms in the
matching coefficients $D_{S^2,s}^{(2)}$ and $c_s$.  We will see that
the RGI in the matching coefficients plays an important role to make
the result more factorization scale independent.  The results obtained
within a strict perturbative expansion (see Ref.~\cite{Penin:2004ay})
are labelled as LO, NLO and NNLO respectively and, after resummation
of logarithms, as LL, NLL and NNLL.  Taking into account the static
potential exactly, using numerical methods as described in the
previous sections, we obtain improved predictions for the relativistic
corrections that we label by including "I" to the previous labelling:
NLLI (including $c_s$ with NLL precision and the improved relativistic
correction $\delta \rho_n$) and NNLLI ($c_s$ with NNLL precision and
the improved relativistic correction $\delta \rho_n$). For comparison
we will also consider the result without resummation of the logarithms
in the matching coefficient, NNLOI ($c_s$ with NNLO precision and the
improved relativistic correction $\delta \rho_n$). For both, NNLLI and
NNLOI we will consider the results taking the RGI static potential at
LO, NLO, NNLO and NNNLO.

From the point of view of a double counting in $\als$ and $v$ the NLL
result (with NLL precision for $c_s$) can be considered as ${\cal
  O}(\als, v^0)$ whereas NLLI is ${\cal O}(\als, v^2)$ and NNLLI is
${\cal O}(\als^2, v^2)$.  As a general trend, moving from NLL to NLLI
improves the scale dependence.  This is due to the fact that, by using
the RGI, NNLO ${\cal O}(\als^2)$ logarithms count as NLL and can be
matched with a part of the scale dependence of the relativistic ${\cal
  O}(v^2)$ correction.  Note as well that the inclusion of $c_s$ with
NNLL precision accounts for ${\cal O}(\als^3)$ leading logarithms and
beyond. Those should be cancelled by the inclusion of the subleading
scale dependence of the relativistic correction. Most of it is
actually built in by the numerical evaluation of the relativistic
correction with the RG potential. In principle, this should be
reflected in an improvement in the scale dependence in going from NLLI
to NNLLI.  On the other hand, this double counting in $\als$ and $v$
scheme produces an unmatched scheme dependence, which can only be
matched by working at the same order in $\als$ and $v$.

We have also studied the dependence on the specific
$r$-renormalization scheme of ${\hat
  B}^{(r)}_{V_s^{(0)}}(E^{(0)}_n;\mu)$. This dependence should vanish
when combined with $c_r^{\MS}$. In particular we have studied the
effect of eliminating $\gamma_E$ in the logarithms in \Eqn{eq:Ar} and
consequently using \Eqn{eq:CDbis} for $c_r^{\MS}$. Note that this is
actually equivalent to using ${\hat B}^{(r)}_{V_s^{(0)}}(E^{(0)}_n;\mu
e^{-\gamma_E})$.  We have checked that (at least in the cases where
the series converges) this dependence fades away when considering the
potential with increasing accuracy. The reason is that the $\gamma_E$
terms that appear at higher orders get more accurately described as we
increase the order of our computation. This increases our confidence
in the perturbative approach. The introduction of $\gamma_E$ in the
scale $\mu$ makes the different terms in the expansion approach the
asymptotic result faster, but the effect is not very significant.

In the following subsections we will consider in turn the cases of 
top, bottom and charm.

\subsection{Top}

We start with the top since it is the cleanest possible case, where we
expect best convergence. The scales are fixed as
$\mu_F=\mu_r=\mu_s=20$~GeV and $\mu_{us}=10$ GeV and we work in the PS
scheme.

\begin{figure}[t]
\epsfxsize=0.8\textwidth
\centerline{\epsffile{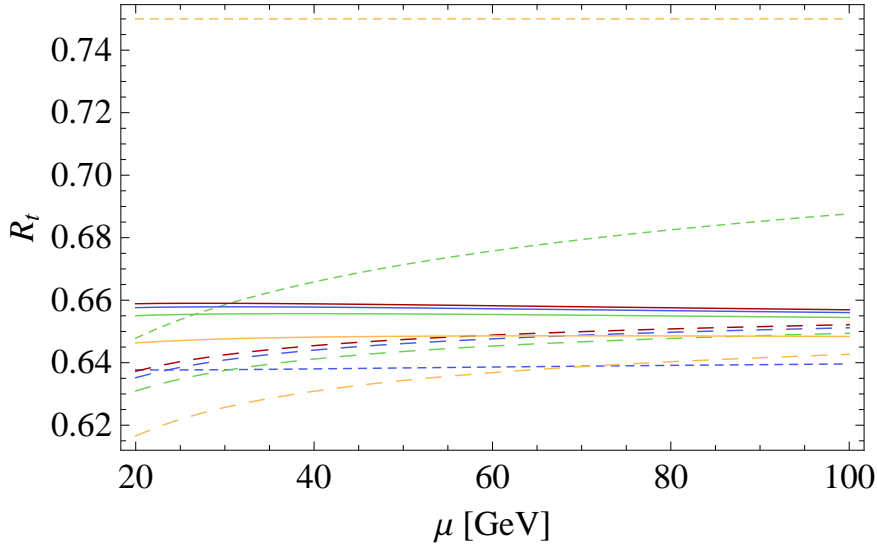}}
\caption{\label{fig:ratioRSp2_t} Decay ratio in the PS scheme at NNLOI
  (dashed) and NNLLI (solid) at different orders in $\als$ in the
  static potential (${\cal O}(\als)$: yellow; ${\cal O}(\als^2)$:
  green; ${\cal O}(\als^3)$: blue; ${\cal O}(\als^4)$: red). For
  reference we also include the LL, NLL, and NNLL results
  (short-dashed).}
\end{figure}

In Figure~\ref{fig:ratioRSp2_t} we show the decay ratio at NNLOI
(dashed lines) and NNLLI (solid lines) at different orders in $\als$
in the static potential (LO: yellow; NLO: green; NNLO: blue; NNNLO:
red). For reference we also include the LL, NLL, and NNLL results
(short-dashed lines) obtained within a strict perturbative expansion.
Comparing the NNLOI with the NNLLI curves, it can be seen that the
inclusion of the RG matching coefficients has a significant impact in
reducing the scale dependence. Also, there is a sizable gap when
moving from NNLL to NNLLI even if we take the LO RGI static potential
(which includes the $r$ running producing the shift we observe in the
plot). The inclusion of subleading corrections to the potential
produces a convergent effect. Actually, the NLO RGI static potential
result is already quite close to the asymptotic result. This may allow
to define a counting in $v$, by taking the asymptotic limit of the
series.  The potential problem is that this counting in $v$ is scheme
dependent.

\begin{figure}[t]
\begin{center}
\epsfxsize=0.8\textwidth
\epsffile{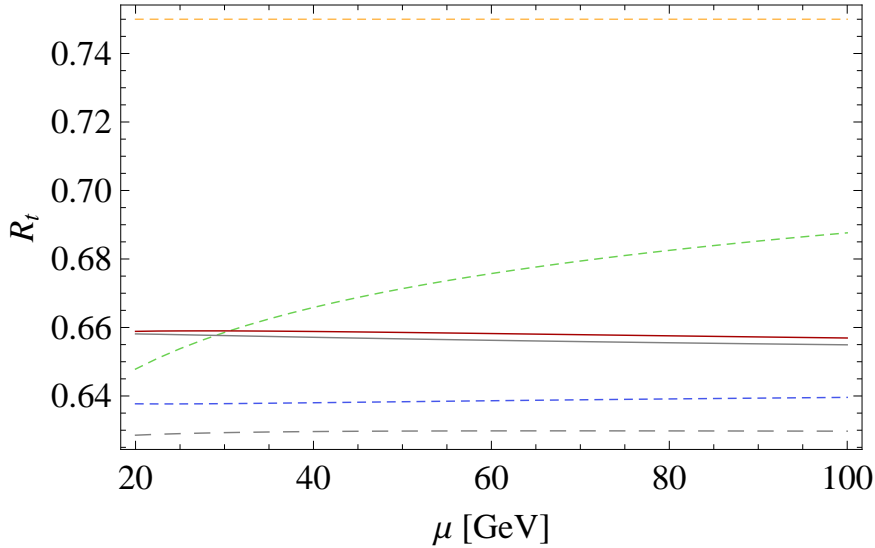}
\end{center}
\caption{\label{fig:decay_top} Decay ratio in the PS scheme at NLLI in
  the $\MS$ (grey dashed) and hard-matching scheme (grey solid) and at
  NNLLI (red solid). For reference we also include the LL, NLL, and
  NNLL results.  }
\end{figure}

To study this scheme dependence, in Figure~\ref{fig:decay_top} we show
the decay ratio at NLLI in the $\MS$ and hard-matching scheme (see
Ref.~\cite{Penin:2004ay} and Eq. (\ref{deltaHMKiyo})) and at NNLLI,
all of them at ${\cal O}(\als^4)$ in the static potential. These
results are compared to the LL, NNL and NNLL results.  Moving from NLL
to NLLI improves the scale dependence no matter what scheme is used.
As we have already discussed, this is due to the fact that, by using
the RG, NNLO ${\cal O}(\als^2)$ logarithms count as NLL and can be
matched with a part of the scale dependence of the relativistic ${\cal
  O}(v^2)$ correction.  On the other hand there is a sizable gap
between the NLLI result obtained in the $\MS$ and hard-matching
scheme. The latter is much closer to the full NNLLI result. The reason
is that the two-loop hard correction is much smaller in the
hard-matching scheme compared with the $\MS$ scheme. This could
indicate that the hard-matching scheme leads to a more convergent
series but it cannot be ruled out that this smallness is accidental
for ${\cal O}(\als^2)$. Therefore, we believe this gap gives a
conservative estimate of the remaining uncertainties. Note that it is
much larger than the other sources of uncertainties considered in this
paper.  For instance, we have also investigated the $\mu_s$ dependence
and observed that it gets smaller when we consider higher orders in
the static potential, pointing to the fact that the long-distance tail
of the potential does not have a significant impact on the
determination of the decay ratio.  A similar comment applies to the
renormalon scheme dependence. Therefore, in summary we find nice
convergence in the top quark case.

\subsection{Bottom}

\begin{figure}[t]
\begin{center}
\epsfxsize=0.8\textwidth
\epsffile{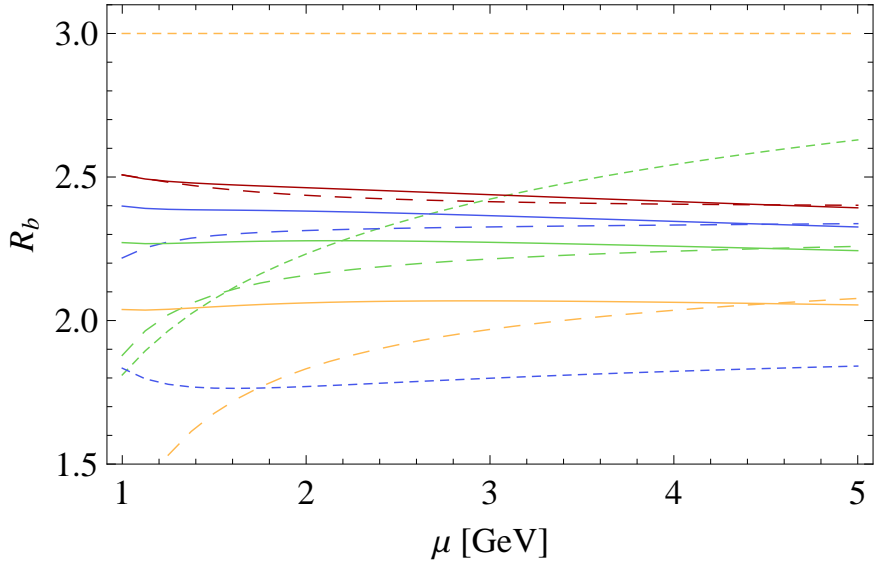}
\end{center}
\caption{\label{fig:ratioRSp2_b} Decay ratio in the PS scheme at NNLOI
  (dashed) and NNLLI (solid) at different orders in $\als$
  in the static potential (${\cal O}(\als)$: yellow; ${\cal
    O}(\als^2)$: green; ${\cal O}(\als^3)$: blue; ${\cal O}(\als^4)$:
  red). For reference we also include the LL, NLL, and NNLL results
  (short-dashed).  }
\end{figure}

Turning to the bottom case, in Figure~\ref{fig:ratioRSp2_b} we show the
decay ratio in the PS scheme at NNLOI and NNLLI at different orders in
$\als$ in the static potential. For reference we also include the LL,
NLL, and NNLL results. We use $\mu_r=\mu_F=\mu_s=2$~GeV and
$\mu_{us}=1$~GeV.  Again we can see that the inclusion of the RG
matching coefficients has a significant impact in reducing the scale
dependence. As in the top case, there is a sizable gap when moving
from NNLL to NNLLI. The bulk of it is already obtained by taken the 
NLO(LO) RGI static potential in the PS(RS') scheme. 
The inclusion of subleading corrections to the potential
produces a smaller effect, yet sizable.  Compared to the top case the
magnitude of the corrections is larger and the convergence using the
static potential at different orders is worse, in particular in going
from the ${\cal O}(\als^3)$ to the ${\cal O}(\als^4)$ approximation of
the static potential. Nevertheless, one can still see a band (though
much wider than for top) where to roughly define a counting in $v$. We
should also stress that using the ${\cal O}(\als^4)$ RGI potential has
some ambiguities, since ultrasoft effects enter at this
order. Therefore, it can not be considered complete.

\begin{figure}[t]
\begin{center}
\epsfxsize=0.8\textwidth
\epsffile{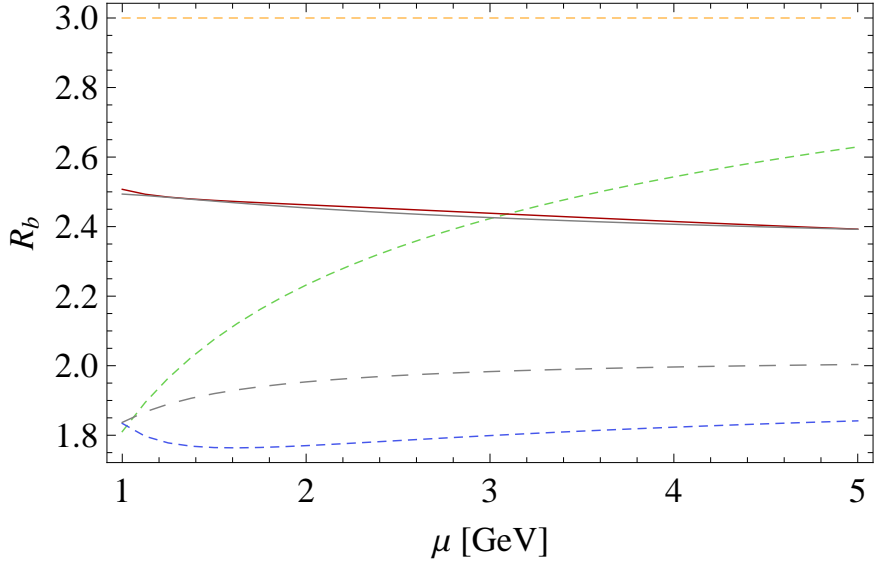}
\end{center}
\caption{\label{fig:decay_bottom} Decay ratio in the PS scheme at NLLI
  in the $\MS$ (grey dashed) and hard-matching scheme (grey solid) and
  at NNLLI (red solid). For reference we also include the LL, NLL, and
  NNLL results.  }
\end{figure}

We study the scheme dependence in Figure~\ref{fig:decay_bottom},
showing the decay ratio at NLLI in the $\MS$ and hard-matching scheme
and at NNLLI. In all cases the static potential is taken at ${\cal
  O}(\als^4)$. These results are compared with the LL, NNL and NNLL
results. The general pattern of the results is similar to the top
case. Moving from NLL to NLLI improves the scale dependence
irrespective of the scheme used.  However, there is a sizable gap
between the NLLI result obtained in the $\MS$ and hard-matching
scheme, the latter is much closer to the full NNLLI result. As for
top, we take this gap for a conservative estimate of the remaining
uncertainty.  Again, this gap is larger than other sources of
uncertainties considered in this paper, like the splitting associated
to different orders in the static potential, the $\mu_s$ or renormalon
scheme dependence. Either way the errors are obviously larger here
than in the top case. In particular we have found a larger sensitivity
to $\mu_s$ and the specific implementation of the initial conditions.

We use this analysis to obtain an updated prediction for
$\Gamma(\eta_b(1S) \rightarrow \gamma\gamma)$.  For the central value
we use the NNLLI result with $\mu=2$~GeV and the set of parameters
described before, obtaining 0.544~keV. The theoretical error has been
estimated considering the difference between the NLLI (in the $\MS$)
and NNLLI result for $\mu=2$~GeV. We obtain 0.146~keV for this
error. As we have already mentioned, we have checked that the
uncertainties due the variation of these parameters, the scheme, or
the consideration of different order in $\als$ in the potential, is
much smaller than the error quoted. Another source of error is
experimental, coming from $\Gamma(\Upsilon(1S) \rightarrow
e^+e^-)=1.340\pm0.018$~keV \cite{Amsler:2008zzb}.  This produces a
very small error: $\pm 0.007$~keV. Finally, we have also computed the
error associated to the indetermination of $\als(M_z)=0.118 \pm
0.003$. This error is even smaller: ${}^{+0.002}_{-0.004}$~keV. We
combine the last two errors in quadrature and add linearly to the
theoretical error (which completely dominates the error).  After
rounding we obtain $\Gamma(\eta_b(1S) \rightarrow \gamma\gamma) = 0.54
\pm 0.15$~keV.

\subsection{Charm}

Finally we consider the charmonium ground state. The applicability of
our weak coupling approach to this system is doubtful. Nevertheless,
we will find it rewarding that the reorganization of the perturbative
expansion significantly improves the agreement with the experimental
data. Again we will use the PS scheme and set $\mu_s=1.5$~GeV,
$\mu_r=1$~GeV and $\mu_F=1=\mu_{us}=0.7$~GeV.

\begin{figure}[t]
\begin{center}
\epsfxsize=0.8\textwidth
\epsffile{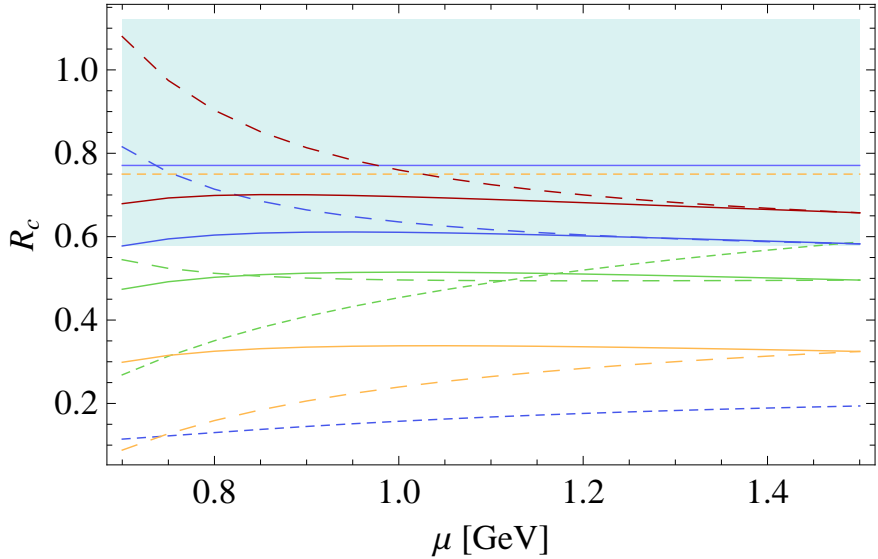}
\end{center}
\caption{\label{fig:ratioRSp1_c} Decay ratio in the PS scheme at NNLOI
  (dashed) and NNLLI (solid) at different orders in $\als$ in the
  static potential (${\cal O}(\als)$: yellow; ${\cal O}(\als^2)$:
  green; ${\cal O}(\als^3)$: blue; ${\cal O}(\als^4)$: red). For
  reference we also include the LL, NLL, and NNLL results
  (short-dashed).  The light blue band represents the experimental
  error of the ratio where the central value is given by the
  horizontal solid line.}
\end{figure}

In Figure~\ref{fig:ratioRSp1_c} we show the decay ratio at NNLOI and
NNLLI at different orders in $\als$ in the static potential. For
reference we also include the LL, NLL, and NNLL results. The
experimental result, using $\Gamma(J/\psi\to e^+ e^-)=5.55\pm
0.14$~keV and
$\Gamma(\eta_c\to\gamma\gamma)=7.2\pm0.7\pm2.0$~keV~\cite{Amsler:2008zzb}
is shown as the light blue band, with the central value indicated by
the horizontal solid light blue line. Once more we can see that the
inclusion of the RG matching coefficients improves the scale
dependence and there is a sizable gap when moving from NNLL to NNLLI.
The inclusion of subleading corrections to the potential produces a
slightly smaller though still quite large effect. Compared to the bottom
case the magnitude of the corrections is larger and the convergence is
worse. We find the same problem in the associated evaluations of the
energy and the wave function at the origin. Despite these
shortcomings, the effect goes in the direction of bringing agreement
with experiment.

\begin{figure}[t]
\begin{center}
\epsfxsize=0.8\textwidth
\epsffile{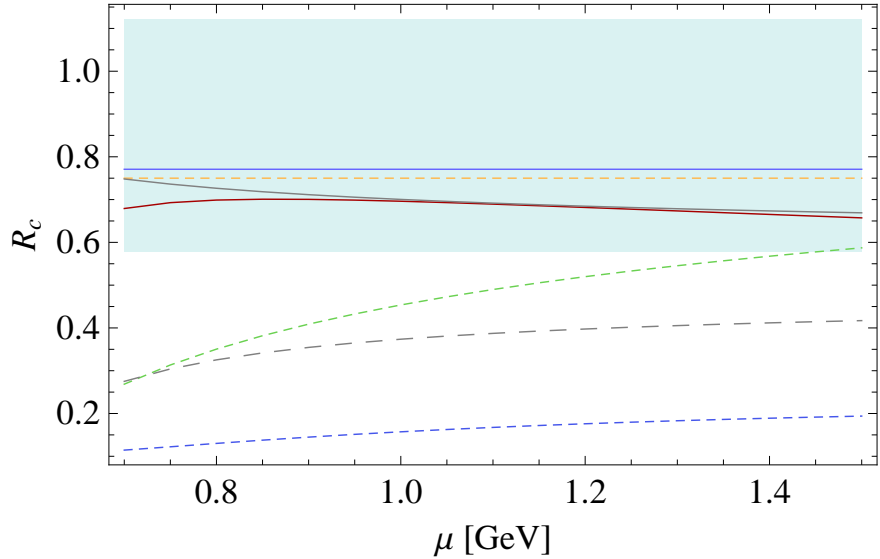}
\end{center}
\caption{\label{fig:decay_charm} Decay ratio in the PS scheme at NLLI
  in the $\MS$ (grey dashed) and hard-matching scheme (grey solid) and
  at NNLLI (red solid).  For reference we also include the LL, NLL,
  and NNLL results and the experimental ratio.}
\end{figure}

We study the scheme dependence by showing the decay ratio at NLLI in
the $\MS$ and hard-matching scheme and at NNLLI in
Figure~\ref{fig:decay_charm}. The static potential is taken at ${\cal
  O}(\als^4)$. The discussion is pretty similar to the top and bottom
case.  Moving from NLL to NLLI improves the scale dependence no matter
what scheme is used.  On the other hand there is a sizable gap between
the NLLI result obtained in the $\MS$ and hard-matching scheme, the
latter being much closer to the full NNLLI result. The reason is the
same as for top and bottom.  Taking this gap for an estimate of the
typical size of the uncertainties produces an error of around 50\% in
the ratio. This encodes most of the experimental band and it is
significantly larger than the typical split produced by working at
different orders in $\als$ in the static potential.

\section{Conclusions}

We have considered a different power counting in potential NRQCD by
incorporating the static potential exactly in the leading order
Hamiltonian. In this scheme we compute the leading relativistic
corrections to the inclusive electromagnetic decay ratios. The effect
of this new power counting is dramatic for charm, large for bottom,
and sizable even for top. In the case of bottom, we produce an updated
value for the $\eta_b$ decay to two photons 
\begin{equation}
\Gamma(\eta_b(1S) \rightarrow \gamma\gamma)=0.54 \pm 0.15 \, {\rm keV}.
\label{GammaEta}
\end{equation} 
In the case of charmonium, this scheme brings consistency between the
weak coupling computation and the experimental value of the decay
ratio, but the theoretical error is large.

It is worth emphasizing that in the case where our expansion is more
reliable, i.e. the top and bottom case, the bulk of the correction
comes from using the first two orders of the RGI potential.  The
effect of higher-order corrections in the RGI potential is relatively
small. The details of the importance of higher-order corrections
depends on the scheme.  In the RS' already the LO RGI potential gives
the bulk of of the correction whereas in the PS two terms in the
expansion are needed.  Irrespectively, they both converge as one goes
to higher orders.

This approach could open the possibility to reorganize the
perturbative series in a controlled way. We stress again that this is
also relevant for top. Therefore, it is not a strong coupling effect
but rather reflects the need of a more optimal resummation of
perturbation theory. This might call for a reanalysis of previous
results in this new scheme. It is an open question whether there is a
similar effect in the case of the hyperfine splitting. We leave this
discussion for a forthcoming paper.

It would be misleading to only assign a theoretical error from the
scale dependence. This is particularly obvious in the charm case,
where the scale dependence by no means reflects a reasonable estimate
of the size of higher-order corrections, which are difficult to
estimate and can only be inferred from the apparent convergence of the
expansion.

Our formalism is flexible enough so that, with little effort, we could
replace the perturbative static potential by any potential, in
particular, by one fitted to non-perturbative lattice data.  This could
be of particular relevance for charmonium but it could also be of help
for bottomonium, provided the static potential is know with enough accuracy 
in the unquenched approximation. This would eliminate the error associated 
to higher order terms in the static potential, but not the error due to 
higher order terms in the hard matching coefficient and the associated RG 
improvement.

\end{document}